\documentclass[conference,compsoc]{IEEEtran}
\usepackage{listings}
\usepackage{tikz}
\usepackage{array}
\usepackage{varwidth}
\usepackage{subcaption}
\usepackage{booktabs}
\usepackage{epigraph}
\usepackage{tablefootnote}
\usepackage{varwidth}
\usepackage{hyperref}
\usepackage{enumitem}
\setitemize{noitemsep,topsep=0pt,parsep=0pt,partopsep=0pt}
\usepackage{threeparttable}

\usepackage[most]{tcolorbox}

\usepackage{etoolbox}

\newlist{compactenum}{enumerate}{4}
\setlist[compactenum,1]{nolistsep,label=(\alph*)}

\renewcommand{\paragraph}[1]{\vspace{5pt}\noindent\textbf{#1.}\xspace}
\newcolumntype{R}{>{\raggedleft\arraybackslash}X}
\newcolumntype{L}{>{\raggedright\arraybackslash}X}
\newcolumntype{P}[1]{>{\centering\arraybackslash}p{#1}}

\newcommand{\eg}{e.g., }

\newcommand{\ie}{i.e., }

\tcbset{textmarker/.style={%
        enhanced,
        parbox=false,boxrule=0.1mm,boxsep=0.1mm,arc=1mm,
        outer arc=1mm,left=2mm,right=1mm,top=2pt,bottom=2pt,
        toptitle=1mm,bottomtitle=0.5mm}
}
\newtcolorbox{importantBox}{textmarker,
    borderline west={4pt}{0pt}{gray},
    colback=orange!10!white}
\newcommand{\important}[1]{\begin{importantBox} #1 \end{importantBox}}
        
\newcommand{\para}[1]{\smallskip\noindent\textbf{{#1.}}}

\iftrue
\newcommand\kurt[1]{\textcolor{blue}{Kurt: #1}}
\newcommand\elie[1]{\textcolor{blue}{Elie: #1}}
\newcommand\todo[1]{\textcolor{red}{#1}}
\else
\newcommand{\kurt}[1]{}
\newcommand{\elie}[1]{}
\newcommand\todo[1]{}
\fi

\makeatletter
\patchcmd{\epigraph}{\@epitext{#1}}{\itshape\@epitext{#1}}{}{}
\makeatother

\begin{document}

\def\FrameWorkName{DroidCCT}

\title{\textsc{\FrameWorkName}: Cryptographic Compliance Test via Trillion-Scale Measurement}

\author{
    \IEEEauthorblockN{
    Daniel Moghimi, 
    Alexandru-Cosmin Mihai,
    Borbala Benko,
    Catherine Vlasov, \\
    Elie Bursztein,
    Kurt Thomas,
    László Siroki,
    Pedro Barbosa,
    Rémi Audebert
    } 
    \IEEEauthorblockA{Google}
}

\pagestyle{plain} 
\maketitle

\begin{abstract}

We develop \FrameWorkName, a distributed test framework to evaluate the scale of a wide range of failures/bugs in cryptography for end users.
\FrameWorkName\ relies on passive analysis of artifacts from the execution of cryptographic operations in the Android ecosystem to identify weak implementations.
We collect trillions of samples from cryptographic operations of Android Keystore on half a billion devices and apply several analysis techniques to evaluate the quality of cryptographic output from these devices and their underlying implementations.
Our study reveals several patterns of bugs and weakness in cryptographic implementations from various manufacturers and chipsets.
We show that the heterogeneous nature of cryptographic implementations results in non-uniform availability and reliability of various cryptographic functions.
More importantly, flaws such as the use of weakly-generated random parameters, and timing side channels may surface across deployments of cryptography.
Our results highlight the importance of fault- and side-channel-resistant cryptography and the ability to transparently and openly test these implementations.

\end{abstract}

\section{Introduction}

Software bugs~\cite{lazar2014does,mouha2018finding,bernstein2013factoring,nemec2017return,breitner2019biased}, hardware failures~\cite{sullivan2022open}, and back doors~\cite{checkoway2016systematic} are just some of the many possible threats to cryptographic implementations. 
If these flaws are not properly mitigated, they can lead to compromised credentials~\cite{ryan2023passive,rowe2023curious,heninger2012mining}, and IP theft~\cite{hettwer2021side}, among others.
For end users and app developers, weak cryptography creates a false sense of privacy and security~\cite{jancar2020minerva,moghimi2020tpm,valenta2018search,adrian2015imperfect,beurdouche2017messy}.
Subtle vulnerabilities such as ROCA~\cite{nemec2017return} or TPM-Fail~\cite{moghimi2020tpm} have shown that obscure cryptographic implementations can degrade end-user security if not implemented with care.

There are several challenges in ensuring reliable cryptographic implementations for user devices, such as those deployed in mobile phones, handheld devices, smart appliances, and cars:
\textbf{(1)} These implementations are typically proprietary and therefore not externally reviewed. 
\textbf{(2)} The diversity of devices makes it expensive for third parties to review various cryptographic stacks.
\textbf{(3)} Some devices are made for low-end markets, such as affordable mobile phones, and are likely not as rigorously scrutinized by professional security teams. 
These challenges leave app developers with hidden risks when relying on these cryptographic stacks.

In this work, we focus on the Android Keystore, which is used by billions of devices connected to the Internet, most of them mobile phones. 
Android Keystore~\cite{aks} was introduced in 2013 (Android 4.3) to improve the protection of cryptographic keys.
It extends the Java KeyStore API with a standard set of functions that have extra protections against malicious apps, compromised operating systems, or even physical attacks.
%. %
Although Android Keystore provides a standard API, the underlying implementations are at the discretion of the device and the chipset manufacturer, which can vary widely in terms of performance and reliability~\cite{keymasterhal}.
As a result, implementations may not be compliant in terms of functionality, performance, or security. 
In this context, we aim to answer the following:
\begin{itemize}
    \item \emph{What are common flaws in these implementations?}
    \item \emph{How often do they surface across different devices?}
    \item \emph{Can these flaws weaken cryptography and risk users?}  
\end{itemize}

To answer these questions, we develop \FrameWorkName, a framework for Android cryptographic compliance testing.
\FrameWorkName\ uses the Android Keystore API that provides an opportunity to test the behavior of a wide range of cryptographic stacks, including hardware, firmware and software.
We develop signals and collect samples from exercising cryptographic functions that are supported by most Keystore implementations.
These samples include random and hard-coded inputs and outputs (e.g., public keys, signatures, encrypted blobs), timing, and errors from exercising these functions.
We integrate signals into the Play Integrity API to collect samples from hundreds of millions of devices and aggregate them based on the unique device model and the execution environment of cryptographic functions.
Finally, we apply several analysis pipelines to check for errors, inconsistencies, weaknesses, and vulnerabilities in implementations solely based on examining these samples. 
Our analysis pipelines include:
\begin{compactenum}
    \item \textbf{Cross-validation:} Check for inconsistency between on-device and independent evaluation of operations.
    \item \textbf{Randomness Testing:} Test random numbers in cryptography to identify weak cryptographic parameters.
    \item \textbf{Public Key Analysis:} Apply black-box public key cryptanalysis to identify weak keys.
    \item \textbf{Side Channel Analysis:} Test a known class of critical timing side channels~\cite{moghimi2020tpm,jancar2020minerva,brumley2011remote}.
\end{compactenum}

We deploy \FrameWorkName\ and collect billions of samples from cryptographic stacks inside Android devices and analyze this massive dataset to identify widespread errors, transient faults, weak cryptographic parameters, and timing side channels.
Our results demonstrate that large-scale datasets can help identify reliability and security gaps in cryptography. 
For the first time, we show that aggregating data samples via telemetry from a large number of diverse devices can help find patterns of weakness and vulnerabilities in a large set of hardware and implementation choices.

As a summary, we contribute the following:
\begin{itemize}
    \item \FrameWorkName, a test framework for Android Keystore Cryptography to identify failures and weaknesses in obscure cryptographic implementations.
    \item An overview of the performance and availability of Keystore in the Android echosystem based on collecting a trillion-scale dataset. 
    \item Quantifying failures, weaknesses, and bug patterns in cryptographic implementations across tens of chipsets and implementation choices that impede secure use of cryptography for a subset of end users.
    \item Recommendations on improving these implementations against the discovered flaws. 
\end{itemize}

\para{Artifact}
You can find a test app that shows how we exercised different cryptographic APIs and a data set that reflects the identified bug patterns at \url{github.com/google/droidcct-paper-artifact}.

\para{Responsible disclosure}
We reported our findings to the major Android chipset makers.
They acknowledged our findings and started an investigation to improve the quality of implementations.
More specifically, 
\begin{itemize}
    \item \emph{Google} confirmed that the weak RSA keys discovered are due to a vulnerability that was already reported and fixed under CVE-2022-20117~\cite{gsacve}.
    \item \emph{Unisoc} confirmed the ECDSA signatures with zeros in their latest TEE implementations as a vulnerability tracked under CVE-2025-31719.
    \item \emph{Mediatek} noted that the discovered problems are on products with TEE implementations implemented by by OEMs or third parties such as MicroTrust (Beanpod) TEE, Samsung TEEgris, or Huawei’s TEE. We did not investigate this further, but have shared our findings with both Samsung and Huawei.
\end{itemize}

\section{Preliminary} \label{sec:back}
In this section, we discuss preliminary knowledge.

\subsection{Android Keystore}
Android Keystore~\cite{aks} supports apps with protected storage for cryptographic keys and hardened implementations of cryptographic algorithms. 
App developers can generate and store keys in an isolated and protected environment, where only that environment can access the keys to execute algorithms (e.g., key generation, encryption, signing).

The Android Keystore API is uniform and supports three different key storage and execution environments across various devices. These are 
\begin{compactenum}
    \item \textbf{Software:} A service software that protects keys from compromised applications.
    \item \textbf{TEE:} A trusted firmware that protects keys from a root adversary, i.e., a compromised Android OS via CPU isolation mechanisms, e.g., TrustZone. 
    \item \textbf{StrongBox:} A Common Criteria certified coprocessor (possibly on the same system on a chip---SoC or a separate SoC) that provides hardware-level isolation against a physical attacker in addition to software adversaries.
\end{compactenum}

The features supported by Android Keystore at different API levels are listed in Table \ref{tab:keystore_features} in the Appendix. 
StrongBox was introduced in Android 9 (API level 28), but there is no guarantee that all devices with this API level support it. 
App developers can check if StrongBox is supported by the device, and enable it when they generate or import a key.
When StrongBox is not used/supported, the underlying implementation may be Software or TEE, whichever is available on the device. 
Developers can enumerate the Android Keystore to identify where the key has been stored.

\subsection{Paranoid Cryptanalysis Library}
In sections~\ref{sec:random} and \ref{sec:paranoid}, we rely on the \emph{Paranoid}~\cite{bbparanoid} library for our analysis, which implements several cryptanalytic tests.
Paranoid supports two types of test:
\begin{compactenum}
    \item \textbf{Randomness Tests:} NIST's randomness test suite~\cite{rukhin2001statistical} and several additional tests aim to evaluate the quality of cryptographic random number generators.
    \item \textbf{Key and Signature Tests:} Known attacks against public keys and signatures and identifies weakness in cryptographic artifacts~\cite{lenstra2012ron,heninger2012mining,bernstein2013factoring,breitner2019biased}.
\end{compactenum}

We apply the randomness test to parameters that we can extract from cryptographic artifacts with a known key (e.g., ECDSA nonce).
For EC and RSA keys that we do not have the private key, we apply the black-box public key tests.

\subsection{Analysis of Cryptographic Artifacts}
Previous work has studied cryptographic artifacts from service providers.
Heninger et al. \cite{heninger2012mining} performed a large-scale study of TLS and SSH revealing RSA keys with insufficient entropy.
Valenta et al. \cite{valenta2018search} studied elliptic curves in TLS, SSH, and IPsec, and discovered that a small percentage of hosts do not properly validate curve parameters.
Breitner et al. \cite{breitner2019biased} studied weak ECDSA signatures in cryptocurrencies and found weak keys.
Sullivan et al. \cite{sullivan2022open} relied on network measurements to study hardware faults in signature implementations that led to RSA private key compromises.
Ryan et al. \cite{ryan2023passive} performed a passive analysis of SSH keys and discovered hundreds of weak samples.

Client-side operations have been analyzed on a small scale in lab settings~\cite{vsvenda2016million,moghimi2020tpm,jancar2020minerva,nemec2017return}.
Recently, Svenda et al. \cite{svenda2024tpmscan} analyzed 78 TPM implementations from six vendors, uncovering several vulnerabilities in TPMs, including nonce and timing leakages in ECC algorithms. 

For the first time, we performed a large-scale analysis of cryptographic implementations of Android devices that are used in mobile phones, handheld devices, and cars.

\section{\FrameWorkName} \label{sec:methods}

\begin{figure*}
\centering
\includegraphics[width=0.73\textwidth]{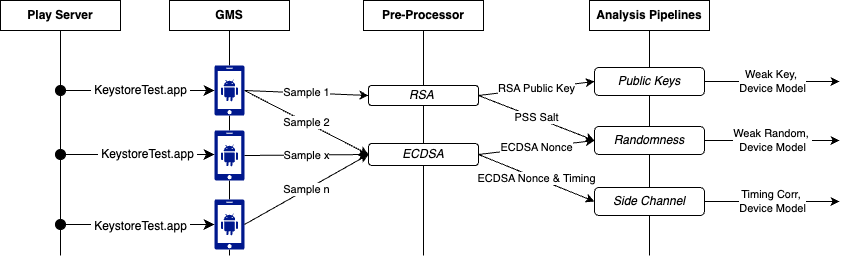}
\caption{\FrameWorkName: A test app that exercises different cryptographic functions collect samples from real devices, then the samples are aggregated and processed for various signals to assess the correctness of implementations. }
\label{fig:droidcct}
\end{figure*}

\FrameWorkName, as depicted in Figure~\ref{fig:droidcct}, relies on a module that executes a suite of cryptographic operations supported by the Android Keystore.
This module is deployed as part of Google Mobile Service (GMS) in Android to collect samples from the implementation of Android Keystore.
After the cryptographic samples are aggregated, we process them and apply several analysis pipelines to evaluate the correctness and performance of the cryptographic implementation for a device model. 
Next, we describe how \FrameWorkName\ collect samples, how we use it to collect billions of samples, our analysis techniques, ethics, and the limitation of our study.

\subsection{Android Module Functionality}
Our test module iterates over a suite of cryptographic operations: signing, encryption, message authentication, and key exchange. 
We limited our selection to features that are available to multiple generations of Android (\ie Android 6.0 --- 14.0)\footnote{Android Keystore was introduced in Android with API level 18, but most of its functions only become available to developers via \texttt{KeyProtection} and \texttt{KeyGenParameterSpec} with API level 23.}  and also different implementation choices (Software, TEE, StrongBox). These functions include:
{Key gen and import} (RSA 1024; EC NIST-P 256; 128-bit Symmetric Key), 
{Signature gen and verify} (RSA-PKCS1; RSA-PSS; ECDSA), 
{Encryption and decryption} (RSA-PKCS1; RSA-OAEP; AES-CBC), 
{Message authentication} (HMAC-SHA256), and  
{Key exchange} (ECDH).

For each test, we either import a fixed key and self-signed certificates (for EC and RSA); or generate a key and certificate using the Android Keystore. We then execute a subset of cryptographic functions supported by that key with a random input. 
For example, a test generates an EC key and then executes both the ECDSA sign and verify operations on a random message. 
If a device supports StrongBox, we attempt to run our test on StrongBox 50\% of the times. 
Otherwise, TEE or Software will run our tests, to the discretion of the implementation.
We log all inputs, outputs\footnote{The output excludes private keys which are not extractable by design.}, execution times, and any errors encountered during execution.

\subsection{Deployment}
In order to gather real-world data, we partnered with Google to deploy our module as part of its Play Integrity API~\cite{playintegrity}. 
This system is designed to ensure the integrity of a device for which cryptographic correctness falls within the scope of checks to which users have already consented.
As part of this integrity check, Google ran our module on a random sample of more than 2.5 billion devices~\cite{android-device-count}. 
Our module was isolated using Android's existing security controls, running as a user application with least privilege. 
The execution was limited to one second to mitigate any impact on the user experience, in line with resources used by existing integrity checks. 
During this time window, our module would run as many tests as feasible and collect variable number of samples depending on the execution speed of the schemes and implementation.

\begin{table}[t]
\centering
\caption{{\bf Dataset overview:} Suites of cryptographic algorithms executed on 600M real-world devices. Each execution consisted of multiple individual operations. Roughly half of tests were executed using an imported key, while the other half used a generated key.}
\resizebox{0.9\linewidth}{!}{
% \begin{tabular}{lllllll}
% \toprule
% Scheme & Count & Key Import & Software & TEE & Strongbox & Faulted \\
% \midrule
% RSA-Sign-PSS & 45.3 billion & 50.81\% & 1.79\% & 95.71\% & 2.50\% & 1.52\% \\
% RSA-Sign-PKCS1 & 47.2 billion & 50.80\% & 1.81\% & 95.69\% & 2.50\% & 1.49\% \\
% RSA-Crypt-PKCS1 & 32.8 billion & 50.80\% & 1.80\% & 95.82\% & 2.38\% & 0.33\% \\
% RSA-Crypt-OAEP & 64.7 billion & 50.72\% & 1.78\% & 95.77\% & 2.45\% & 0.38\% \\
% HMAC-Sign & 26.8 billion & 39.88\% & 1.91\% & 98.03\% & 0.07\% & 1.56\% \\
% EC-Sign & 214.9 billion & 51.82\% & 1.16\% & 95.75\% & 3.09\% & 0.07\% \\
% EC-Exchange & 22.9 billion & 53.96\% & 45.09\% & 53.20\% & 1.72\% & 0.02\% \\
% AES-Crypt & 39.6 billion & 40.36\% & 2.15\% & 97.78\% & 0.07\% & 0.74\% \\
% \bottomrule
% \end{tabular}

% \begin{tabular}{llllll}
% \toprule
% Scheme & Count & Key Import & Software & TEE & Strongbox \\
% \midrule
% RSA-Sign-PSS & 44.2 billion & 50.84\% & 1.54\% & 95.94\% & 2.52\% \\
% RSA-Sign-PKCS1 & 46.1 billion & 50.83\% & 1.56\% & 95.92\% & 2.52\% \\
% RSA-Crypt-PKCS1 & 32.4 billion & 50.84\% & 1.54\% & 96.05\% & 2.40\% \\
% RSA-Crypt-OAEP & 63.9 billion & 50.75\% & 1.53\% & 96.00\% & 2.47\% \\
% HMAC-Sign & 26.3 billion & 39.92\% & 1.81\% & 98.12\% & 0.07\% \\
% EC-Sign & 213.1 billion & 51.87\% & 0.97\% & 95.92\% & 3.11\% \\
% EC-Exchange & 22.9 billion & 54.01\% & 45.07\% & 53.22\% & 1.72\% \\
% AES-Crypt & 39.1 billion & 40.40\% & 2.05\% & 97.88\% & 0.07\% \\
% \bottomrule
% \end{tabular}

\begin{tabular}{llrr}
\toprule
Cryptographic suite & Variant & Count & Imported Key \\
\midrule
RSA Sign & PKCS1 & 47.6B & 50.8\% \\
RSA Sign & PSS & 45.7B & 50.9\% \\
RSA Crypt & OAEP & 65.2B & 50.8\% \\
RSA Crypt & PKCS1 & 33.1B & 50.9\% \\
EC Exchange & -- & 22.8B & 54.1\% \\
EC Sign  & -- & 215.4B & 51.9\% \\
HMAC Sign & SHA256 & 27.1B & 39.6\% \\
AES Crypt & CBC & 40.1B & 40.1\% \\
\bottomrule
\end{tabular}
}
\label{tab:sample_stats}
\end{table}

\subsection{Dataset}
We gathered data over a two-month period from 5 March to 6 May 2024. 
The implementations we analyzed have been in production and use for a decade, which highlights the long-standing nature of issues in cryptographic implementations.
Our data set consists of 1.5 trillion cryptographic samples from roughly 600 million unique devices. We present a breakdown of the suite of tests in our dataset in Table~\ref{tab:sample_stats}, where each suite represents multiple individual tests (\eg import, sign, verify; or generate, encrypt, decrypt). 

In addition to cryptographic and timing data, we also obtained coarse device metadata for the device associated with each test. This metadata includes the form factor (\eg phone, television), manufacturer (\eg Pixel), model (\eg Version 8), manufacturing year, chipset, and the supported API level of the device. 
Based on the DRAM capacity of the device, we also attribute the signal to determine whether the device was considered ``Premium tier'' ($\ge 12$~GB memory), ``Medium tier'' ($5-12$~GB memory), or ``Low tier'' ($<5$~GB memory). 
We use these metadata to analyze trends between chipsets, API levels, and device tiers.

\subsection{Analysis Pipelines}
\FrameWorkName\ groups samples from a unique device and model and applies various analyses.
For the latter, the device fingerprint is a unique identifier of the hardware and software stack for all identical devices.
To identify trends, we group our findings based on the unique chipset, API level, and the computation backing of the cryptographic artifacts.

We implemented several automated data processing and cryptanalysis pipelines using Python and Apache Beam~\cite{beam}.
We used Python \emph{pyca/cryptography} and \emph{gmpy2} to verify cryptographic artifacts (Section~\ref{sec:functional}) and extract private parameters from various cryptographic operations (Section~\ref{sec:random} \& \ref{sec:ecdsatiming}).
We used Paranoid~\cite{bbparanoid} for the randomness test and public key cryptanalysis (Section ~\ref{sec:random} \& \ref{sec:paranoid}).

\subsection{Ethics}%
Collecting data about the behavior of apps, the platform, or the hardware is in accordance with Google Play services to detect \emph{potentially harmful applications (PHAs)}~\cite{PHAs}.
Our module does not impact the user experience and does not introduce significant computational overhead on the user device.
We limited its execution to a maximum of 1 second, and it only runs on a small fraction of randomly connected devices.
It is possible for a device to execute our signal more than once a day depending on whether some apps request device integrity attestation.

In our analysis, we discard samples that come from less than a hundred samples per unique device model to remove the chance of identifying a user based on the samples.

We are reporting our findings to vendors so that they can improve the quality of their implementation.

\subsection{Limitations}
Android keystore does not allow us to extract private keys and perform precise timing measurement.
Our analysis is limited to public keys, signatures, and encrypted blobs.
We can only analyze some of the internal parameters of cryptographic artifacts when we import hard-coded keys.
For example, we recover the ECDSA nonce for those signatures that were generated with hardcoded key, but we cannot recover any parameter used for side-channel mitigation, e.g. masking.
We measure the timing of operations in milliseconds on the API interface.
The underlying Android keystore API introduces measurement noise. %
Consequently, we can use it to derive high-level conclusions about the performance of operations and perform basic timing side-channel analysis.
Therefore, we cannot identify all side-channel vulnerabilities, e.g., those that can be exploited via co-located adversaries exploiting cache timing.

\section{Keystore Ecosystem} \label{sec:perf} \label{sec:overview}
In this section, we give a high-level overview of the Android Keystore ecosystem, including general support for StrongBox and TEE, the aggregate performance of implementations, and the failure rates of cryptographic functions.

\subsection{StrongBox \& TEE support}

\begin{figure}[h]
    \centering
    \includegraphics[width=0.83\columnwidth]{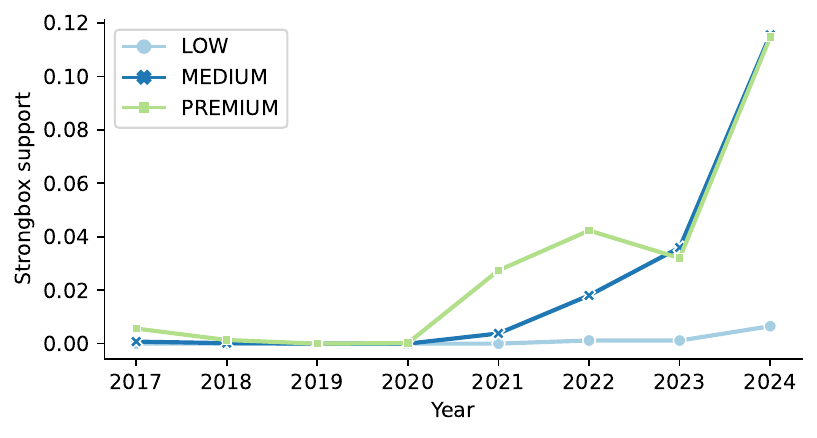}
    \caption{Devices that support strongbox, based on the year they were released, broken down by device tier.}
    \label{fig:strongbox_timeline}
\end{figure}

Across our random sample of devices, 8.7\% of Premium-tier,
7.2\% of Medium-tier, and 0.2\% of Low-tier devices supported StrongBox. 
We present a breakdown of this support based on when a device was first manufactured in Figure~\ref{fig:strongbox_timeline}. 
StrongBox support first appeared in 2021 for Premium-tier devices, and has since accelerated in use such that 11.5\% of Premium- and Medium-tier devices support StrongBox in 2024. 
Of devices without StrongBox, 97.1\% of Premium-tier, 99.2\% of Medium-tier, and 92.7\% of Low-tier devices supported TEE.

\important{\textbf{Availability (1):} TEE is supported on most devices \& StrongBox is growing in premium and medium tier.}

\subsection{Execution Speed}

We estimate the performance of individual cryptographic operations for StrongBox and TEE using the collected execution times, broken down by device tier. We reduce the impact of outliers on our measurements in two ways: (1) we report 90\% percentiles (P90) rather than averages, and (2) as a single device can execute hundreds of operations within our capped runtime of one second (see Section~\ref{sec:methods}), we randomly choose one operation per device, per day, per configuration (\eg import vs. generate, padding choice), while omitting all other samples from our performance analysis. For timing, we use ``wall time'' which includes any delays due to I/O or other resource contention.

\begin{figure}
     \centering
     \begin{subfigure}[t]{0.81\linewidth}
         \centering
         \includegraphics[width=\textwidth]{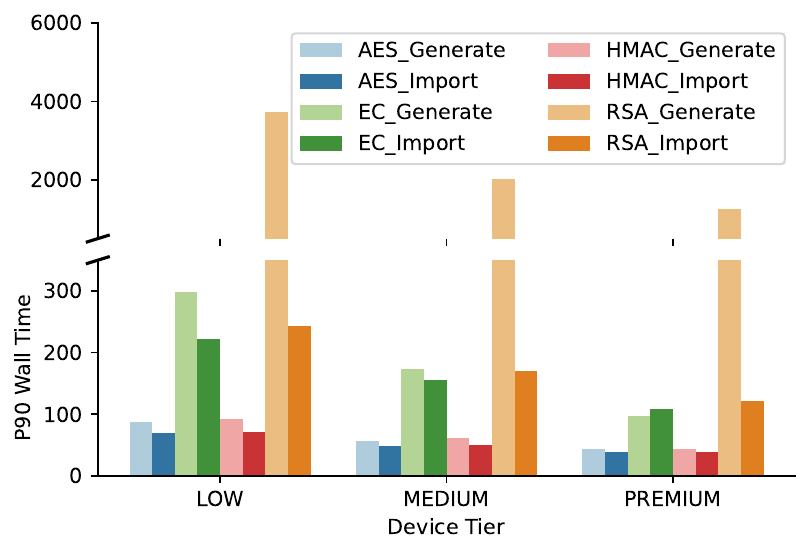}
         \caption{Keys (TEE).}
         \label{fig:time_per_tier_key_tee}
     \end{subfigure}
     \hfill
     \begin{subfigure}[t]{0.81\linewidth}
         \centering
         \includegraphics[width=\textwidth]{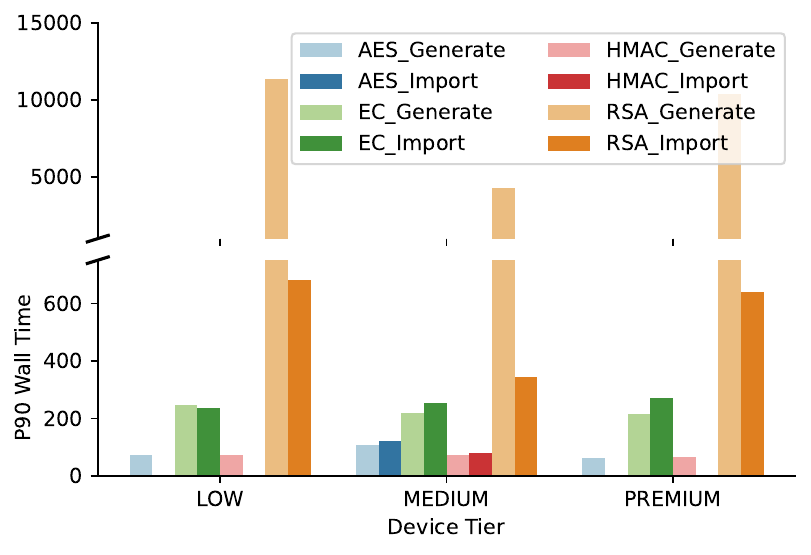}
         \caption{Keys (StrongBox).}
         \label{fig:time_per_tier_key_sb}
     \end{subfigure}
      \hfill
     \begin{subfigure}[t]{0.81\linewidth}
         \centering
         \includegraphics[width=\textwidth]{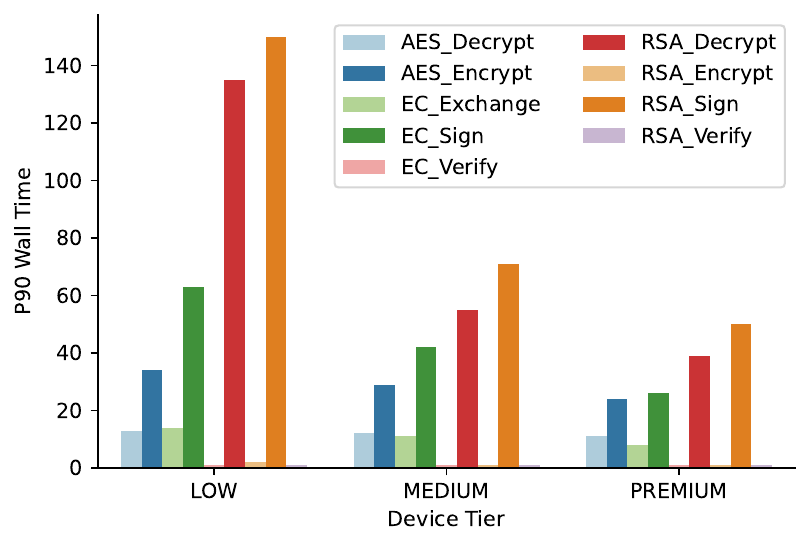}
         \caption{Ciphers (TEE).}
         \label{fig:time_per_tier_op_tee}
     \end{subfigure}
      \hfill
     \begin{subfigure}[t]{0.81\linewidth}
         \centering
         \includegraphics[width=\textwidth]{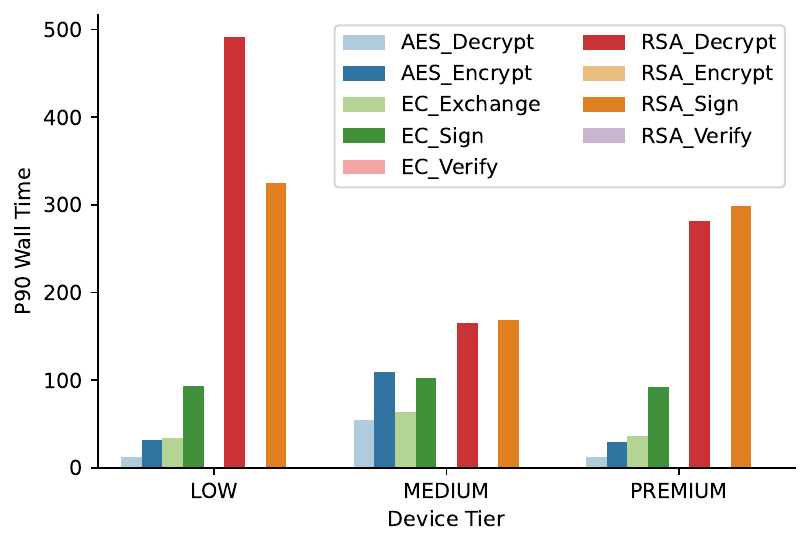}
         \caption{Ciphers (StrongBox).}
         \label{fig:time_per_tier_op_sb}
     \end{subfigure}
    \caption{Execution time of cryptographic operations for StrongBox and TEEs, broken down by device tier and operation type. All times are in milliseconds.}
    \label{fig:execution_time_per_device_type_and_op}
\end{figure}

We observe that for TEE implementations that are typically executed on the application processors, operations are faster on higher end devices, e.g., ciphers on TEEs run 1--3.46X faster on Premium-tier devices versus Low-tier devices (Figure~\ref{fig:execution_time_per_device_type_and_op}).
But that is not the case for StrongBox implementations, computation time is 0.91--1.74X compared to low tier devices.

\important{\textbf{Availability (2):} Cryptography is faster on higher end devices (e.g., TEE running on app processor), but that is not always true for StrongBox based on hardware.}

\subsection{Overall Failure Rate} \label{sec:errors}
We quantify the rate for when a cryptographic operation fails and throws an exception.
Table~\ref{tab:error_stats} shows that the raw error rate ($Rate_{raw}$) for some operations is very high.
We will later confirm that this high rate is due to implementation bugs, and not uniform across devices.
To exclude samples from potentially buggy implementations, we grouped samples based on their chipset model, API level, and error messages, and calculated the \textit{median} of error rates across these groups.
Then, we used this number as a threshold to exclude groups that have an unusually high error rate.
After excluding these sample groups, the error rate for the remaining samples is less than 0.02\% for key import and generation, and less than 0.001\% for other cipher operations, except ECDH, discussed in Section \ref{sec:cipher_errors}.

Since Android Keystore does not raise an exception for signature verification, we cannot calculate its error rate, but we will examine the rate of signature verification failures (when a verification returns \texttt{false}) in Section~\ref{sec:functional}.

\begin{table}
\centering
\caption{Failure rates of cryptographic operations.}
\resizebox{\linewidth}{!}{
\begin{tabular}{lllllll}
\toprule
Algorithm & Variant & Total & Error Count & $Rate_{raw}$ & Median & $Rate_{threshold}$ \\
\midrule
% RSA Generate & - & 503.0M & 500,000 & 0.0994038\% & 0.5642689\% & 0.0096185\% \\
% RSA Import & - & 507.0M & 1.7M & 0.3287982\% & 1.4474481\% & 0.0196619\% \\
% EC Generate & - & 524.1M & 528,700 & 0.1008725\% & 0.2063146\% & 0.0053826\% \\
% EC Import & - & 507.0M & 4.8M & 0.8346968\% & 0.6493506\% & 0.0123023\% \\
% RSA Sign & PKCS1 & 32.4B & 16.8M & 0.0519099\% & 0.1604460\% & 0.0010022\% \\
% RSA Verify & PKCS1 & 32.4B & 101,000 & 0.0003117\% & 0.1004520\% & 0.0000225\% \\
% RSA Sign & PSS & 44.2B & 16.7M & 0.0378127\% & 0.1768389\% & 0.0007858\% \\
% RSA Verify & PSS & 44.2B & 72.3M & 0.1636755\% & 6.1732746\% & 0.0001815\% \\
% ECDSA Sign & - & 213.1B & 93.2M & 0.0437374\% & 0.0287094\% & 0.0001878\% \\
% ECDSA Verify & - & 213.1B & 422,000 & 0.0001980\% & 0.0247792\% & 0.0000690\% \\
% RSA Encrypt & PKCS1 & 32.4B & 14,600 & 0.0000450\% & 0.0350986\% & 0.0000003\% \\
% RSA Decrypt & PKCS1 & 32.4B & 925,900 & 0.0028577\% & 0.0264667\% & 0.0001751\% \\
% RSA Encrypt & OAEP & 63.9B & 30,700 & 0.0000480\% & 0.0562630\% & 0.0000017\% \\
% RSA Decrypt & OAEP & 63.9B & 1.8M & 0.0027464\% & 0.0142279\% & 0.0001442\% \\
% AES Encrypt & - & 39.1B & 13.6M & 0.0348997\% & 0.6294872\% & 0.0007037\% \\
% AES Decrypt & - & 39.1B & 8,300 & 0.0000213\% & 0.0018721\% & 0.0000031\% \\
% HMAC & - & 26.3B & 3.9M & 0.0149803\% & 0.7072948\% & 0.0004499\% \\
% ECDH & - & 22.9B & 7.0B & 30.7292514\% & 99.9576864\% & 14.8093227\% \\
RSA Generate & -     & 503.0M & 500,000 & 0.1\%    & 0.56\% & 0.1e-2\% \\
RSA Import   & -     & 507.0M & 1.7M    & 0.3\%    & 1.45\% & 0.2e-1\% \\
EC Generate  & -     & 524.1M & 528,700 & 0.1\%    & 0.21\% & 0.5e-2\% \\
EC Import    & -     & 507.0M & 4.8M    & 0.8\%    & 0.65\% & 0.1e-1\% \\
RSA Sign     & PKCS1 & 32.4B  & 16.8M   & 0.5e-1\% & 0.16\% & 0.1e-2\% \\
RSA Verify   & PKCS1 & 32.4B  & 101,000 & 0.3e-3\% & 0.10\% & 0.2e-4\% \\
RSA Sign     & PSS   & 44.2B  & 16.7M   & 0.4e-1\% & 0.18\% & 0.8e-3\% \\
RSA Verify   & PSS   & 44.2B  & 72.3M   & 0.1\%    & 6.17\% & 0.2e-3\% \\
ECDSA Sign   & -     & 213.1B & 93.2M   & 0.4e-1\% & 0.03\% & 0.2e-3\% \\
ECDSA Verify & -     & 213.1B & 422,000 & 0.2e-3\% & 0.02\% & 0.7e-4\% \\
RSA Encrypt  & PKCS1 & 32.4B  & 14,600  & 0.4e-4\% & 0.03\% & 0.3e-6\% \\
RSA Decrypt  & PKCS1 & 32.4B  & 925,900 & 0.3e-2\% & 0.02\% & 0.2e-3\% \\
RSA Encrypt  & OAEP  & 63.9B  & 30,700  & 0.5e-4\% & 0.06\% & 0.2e-5\% \\
RSA Decrypt  & OAEP  & 63.9B  & 1.8M    & 0.3e-2\% & 0.01\% & 0.1e-3\% \\
AES Encrypt  & -     & 39.1B  & 13.6M   & 0.3e-1\% & 0.63\% & 0.7e-3\% \\
AES Decrypt  & -     & 39.1B  & 8,300   & 0.2e-4\% & 0.002\% & 0.3e-4\% \\
HMAC         & -     & 26.3B  & 3.9M    & 0.1e-1\% & 0.71\% & 0.4e-3\% \\
ECDH         & -     & 22.9B  & 7.0B    & 30.73\%  & 99.96\% & 14.81\% \\
\bottomrule
\end{tabular}

}
\label{tab:error_stats}
\end{table}

\important{\textbf{Availability (3):} Key and cipher failure rates are less than 0.02\% and 0.001\%, respectively, but the error rate is higher for some devices due to implementation bugs.}

\subsection{Significant Key Errors}
We identified 120 unique error messages due to key operations. 
Some of them, in Figure \ref{fig:err_list_key}, occur with high frequency with the most notable \textsc{Unsupported protection parameter} and \emph{Key size not available} accounting for 35\% and 32\% of the errors, respectively.
By closely looking at the occurrence of similar errors across different device models, we confidently identified two bug patterns:

\begin{figure}
     \centering
         \centering
         \includegraphics[width=\linewidth]{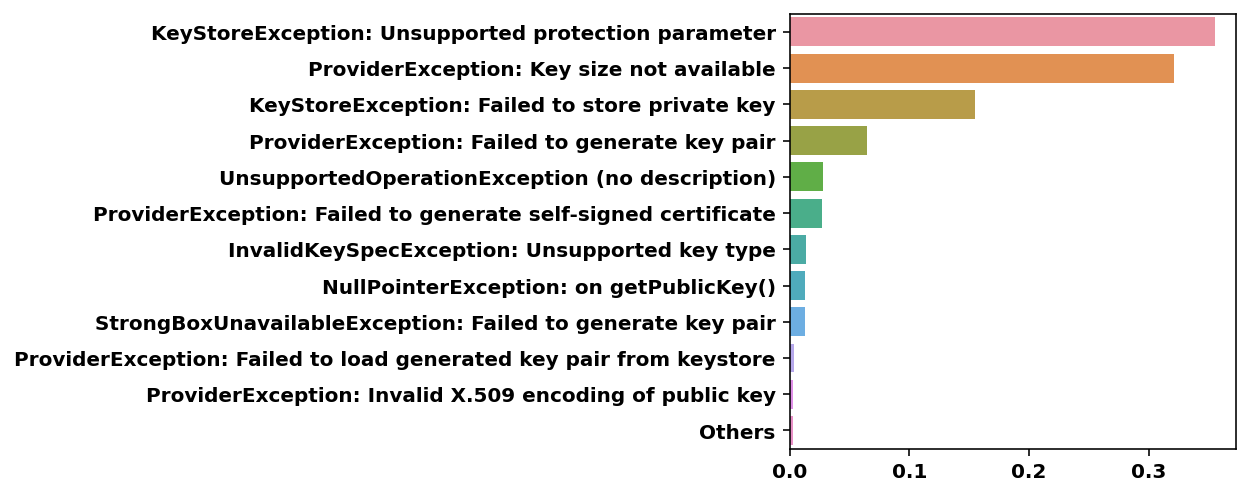}
         \caption{Frequency of errors for key operations.}
         \label{fig:err_list_key}
\end{figure}
\para{StrongBox key import failure}
Importing EC keys to 7 chipsets from Google and Qualcomm fails 50\% of the time (median, minimum, and maximum).
These errors are observed in premium devices with API levels 31-34.
In \FrameWorkName, when StrongBox is available, we use it with probability 0.5.
The 50\% error rate suggests that the import of EC keys to StrongBox for some chipsets always fails, leaving this functionality unusable on a number of premium devices.
The usual error messages for these failures are \emph{Unsupported key size}, \emph{key size not supported}, and the generic \emph{Failed to store private key}.

\para{Unsupported protection parameter}
Importing RSA and EC keys fails with the \texttt{unsupported protection parameter} exception on 41 chipsets from Qualcomm, NVidia, HiSilicon, Samsung, Mediatek and Google.
The error rate is 2.38\%-100.00\% with a median of 88.01\%.
The low minimum error rate suggests that this bug may have already been fixed on some devices, but the 100\% maximum suggests that some chipsets fail to import EC/RSA keys consistently, probably due to an incompatibility bug, where the underlying implementation does not support parameters defined by the keystore API.
\important{\textbf{Bug (1):} Importing EC/RSA keys into StrongBox fails on a number of recent device models, suggesting that the underlying implementation is incompatible or incomplete.}

\subsection{Significant Cipher Errors} \label{sec:cipher_errors}
We investigated the errors generated during cryptographic operations and identified that most of them are due to an \texttt{InvalidKeyException} (Figure~\ref{fig:err_list_other}).
This highlights that even if key import and generation appear to be successful, implementations could suffer from other implementation bugs that make the imported or generated key unusable later.
A complete list of these errors across different chipset groups and API levels is given in Table~\ref{tab:operation_error} in the Appendix.
We investigated the root cause for some of the high error rates:

\begin{figure}
    \centering
    \includegraphics[width=0.7\linewidth]{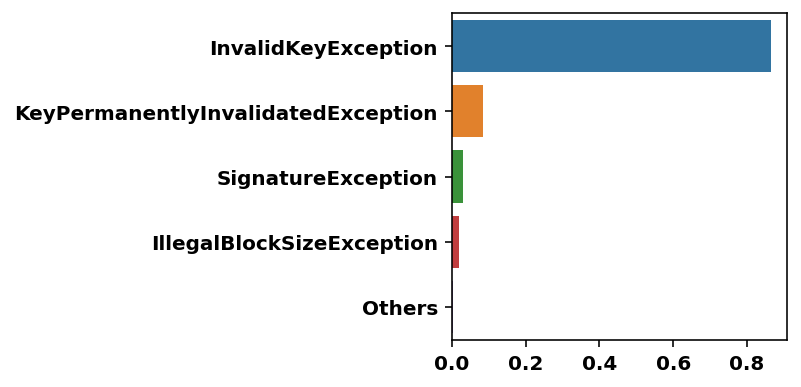}
    \caption{Frequency of different errors among ciphers.}
    \label{fig:err_list_other}
\end{figure}

\para{Invalid blob}
We identified 1,300 devices from 20 chipsets from Unisoc, Qualcomm, Samsung, and Mediatek with TEE and API level 28-29 with a median failure rate of 83.77\%  (min: 0.87\%, max: 100.00\%) that raise an exception with message \emph{invalid key blob} for RSA/ECDSA signing operations.
Similarly, we identified 1,200 devices, five Qualcomm chipsets with TEE implementation, and API level 26-29 with a median failure rate of 99.7\% (min: 74.35\%, max: 100.00\%) with the same error message for HMAC.
These errors occur after both key import and key generation, suggesting that the device has failed to store the key correctly, regardless of the key operation.

\para{StrongBox RSA decryption fails} 
We identified 294 devices with a Qualcomm chipset and API level 33 that have a high rate of RSA decryption failures with \emph{IllegalBlockSizeException} (min: 39.02\%, max: 58.33\%, median: 52.66\%).

\para{ECDH Incompatible purpose}
We identified that 2.6 billion failed ECDH samples throw an exception with the error message \emph{Incompatible purpose}.
This error affects Software/TEE implementations of millions of devices with API level 31-33, and hundreds of chipsets from Unisoc, AMD, Qualcomm, Intel, Samsung, and Mediatek manufacturers.
This high error rate suggests that the underlying implementation does not support \texttt{PURPOSE\_AGREE\_KEY} despite it should be supported according to the API level.

\important{\textbf{Bug (2):} Key storage failure causes signature generation (RSA/ECDSA/HMAC) to fail, which was potentially fixed in later API level 30.

\noindent\textbf{Bug (3):}
RSA decryption fails due to a bug in StrongBox causes invalid padding in a chipset with API level 33.

\noindent\textbf{Bug (4):}
ECDH has not been implemented despite the API level claims to support it across
major manufacturers.}

\section{Functional Cryptanalysis} \label{sec:functional}
Next, we independently verify whether the input and output of a cryptographic function match their expected evaluation. 
We have implemented a data-processing pipeline that processes cryptographic artifacts against their ground truth, which varies depending on the scheme.
In particular, we perform the following analysis.
\begin{itemize}
    \item \textbf{Public Keys:} Check if the exported public key matches the imported private key. 
    \item \textbf{Signatures:} checks if a signature verifies on the device and externally using the public key.
    \item \textbf{Encryption Blobs:} Checks if the input and decryption output match on the device and externally with the hardcoded private/secret key.
    \item \textbf{Exchange/MAC:} Recalculate the shared secret and MAC and check if they match the one generated by the device with hardcoded secret key.
\end{itemize}

\subsection{Invalid Signature}

We previously identified that $0.2e$-$3\%$ of ECDSA, $0.3e$-$3\%$ of RSAPKCS1 signatures, and $0.1\%$ of the RSAPSS signatures cannot be verified on the device (Table~\ref{tab:error_stats}).
Most of them could also not be verified externally due to three major bug patterns:

\para{PSS signatures with invalid salt} %
We used the imported RSA key to recover the PSS salt for 57.3 million failed signatures and identified that, except 235 of them, they all have an invalid salt length.
According to the specification \cite{rsarfc}, we expect the PSS salt to be roughly 32 bytes, based on the SHA-256 output size.
But we observed that the recovered salt is either too small (less than 20 bytes), or too large (more than 220 bytes).
This bug was observed in 93 chipsets from Unisoc, Qualcomm, Intel, MStar, Amlogic, HiSilicon, Samsung, TI, Broadcom and Mediatek at API level 23.
And five chipsets from Qualcomm with API levels 25-29 have produced outputs with the same behavior.

\para{PKCS1 Signatures with incorrect key}
We identified that 97\% of the 112.9 thousand invalid PKCS1 signatures are produced with keys that are generated on-device.
Except for 1.7 thousands, they come from the same measurement, that is, when a measurement generated an invalid key, all samples using that key produced invalid signatures.
Strangely, looking at some of these signatures, they have a similar data pattern as ASN1-encoded ECDSA signatures.
This suggests a bug that caused the PKCS1 signatures to be calculated using an incorrect key, potentially generated or imported by a prior operation.
This bug manifested on several low-end chipsets from HiSilicon, Mediatek, Qualcomm, and Samsung with API levels 23-26, but occasionally appeared in recent API levels.

\para{ECDSA signatures with zero chunks}
We identified that 64\% of the 468.2 thousand invalid ECDSA signatures have contiguous blocks of zero bytes.
We successfully extracted the $(r, s)$ pairs from the ASN1 encoded messages, but the extracted parameters have contiguous chunks of zero bytes up to 28 bytes and a multiple of 4 bytes.
This suggests a bug that causes invalid values to be copied as the output signature.
This pattern was observed on 16 chipsets, including premium devices, from Unisoc with API levels 28-34.

\important{
\textbf{Bug (5):} Invalid PSS implementation affects earliest version of the Android Keystore, API level 23.

\noindent\textbf{Bug (6):} RSAPKCS1 produces signatures with a wrong key that affect older API levels 23-26.

\noindent\textbf{Bug (7):} ECDSA signatures with zero bytes affect recent API levels 28-34 in chipsets from one manufacturer.
}
\subsection{Verification Failures}
We used the public key to audit signature verifications on device.
We identified signatures that verified externally despite failing to verify on the device, but we did not see any trends.
Then, we looked at signatures that failed to verify externally with the public key, despite the successful on-device verification.

We identified 64 RSAPKCS1, 633,200 RSAPSS, and 523 ECDSA signatures that could not be externally verified.
For RSAPSS, 98\% of the samples are due to API level 23, which is related to the previously observed bug due to an invalid PSS implementation in an earlier version of Keystore.
This finding further highlights that there has been an API inconsistency issue that affected API level 23 for both RSAPSS signing and verification, e.g., using a different hash function than specified in the API.

For the remaining unverifiable signatures across all schemes, almost all of them are from low end devices from Amlogic, MStar, Mediatek, and Qualcomm. 
The spread of externally unverifiable signatures for these devices implies that on-device signature verification may succeed even for invalid signatures potentially due to accidental faults~\cite{hochschild2021cores}.

\important{
\textbf{Integrity (1):} Falsely passing signature verification for invalid signatures potentially due to accidental hardware failures affects a very small number of low-end devices. 
}

\subsection{Mismatching Encryption}
Next, we looked at cases where the input of an encryption operation does not match the decrypted output on-device or externally.
We identified 2,435 RSA and 30 AES encryption/decryption pairs that do not match on the device.
We identified the following two trends:

\para{Invalid RSA output}
We identified that 71\% of the cases where decrypted messages have no correlation with the RSA input (OAEP and PKCS1) are from 6 chipsets from Intel with API levels 23-25.

\para{Invalid RSA output size}
We identified that another 27\% of mismatching RSA decryptions are related to OAEP on 15 Mediatek chipsets with API levels 24, and 25.
For these samples, the size of the decrypted message is 254-256 bytes, which does not match the expected 32-byte input that we requested to encrypt with the API.
Among these samples, we identified 583 with all-zero decrypted output.

\important{
\textbf{Bug (8):} Mismatching RSA functions affect one manufacturer with older API levels 23-25.

\noindent\textbf{Bug (9):} Buggy OAEP encryption affects one manufacturer with older API level 24-25.
}

\subsection{Mismatching HMAC}
Next, we looked at HMAC and compared the two MACs that are expected to match for the same message.
We identified 1,938 samples where the two outputs do not match.

\para{HMAC using invalid key}
We identified that in 98\% of the cases where the two HMAC operations do not match, the output is not the expected 32 bytes and close to 72 bytes.
When we looked closely, we identified that the outputs have the same pattern as an ASN1 encoded ECDSA signature.
This behavior for HMAC is similar to the buggy pattern described earlier where the wrong key is used to calculate the RSA signature.
This bug affects three Mediatek and Qualcomm chipsets with API level 29.
We also compared the MAC for the samples with the static key, and it revealed the same problem where the externally calculated MAC does not match the on-device MAC, and the on-device MAC looks like an ECDSA signature.
\important{\textbf{Bug (10):} HMAC producing signatures with a wrong key affects two chipsets  with API level 29.}

\subsection{Silent Data Corruption} \label{sec:sdc}
We attribute low-rate failures to transient errors and silent data corruptions~\cite{hochschild2021cores}.
Such errors may have consequences and weaken cryptography~\cite{ryan2023passive}.

For invalid signatures, after excluding errors that are potentially due to bugs, the rate of invalid signatures due to otherwise transient errors is lower, as shown in Table~\ref{tab:sign_verify_stats_excluding_high}.
We still couldn't reason about hundreds of thousands of invalid ECDSA signatures, which may be due to other less common implementation bugs, affecting older API levels.
For RSA signatures, after excluding buggy samples, we see a failure after 10 billion signing operations, suggesting that signature failures due to transient errors such as silent data corruption (SDC) are rare~\cite{hochschild2021cores}.
Similarly, on-device verification fails for correct signatures after 25 million attempts (Table~\ref{tab:sign_sdc}).

\begin{table}
\centering
\caption{Rate of invalid signatures after excluding the ones due to trendy buggy patterns.}
\resizebox{0.8\linewidth}{!}{
\begin{tabular}{llll}
\toprule
Algorithm & Failed Rate & Failed Median & Failed $Rate_t$ \\
\midrule
RSAPSS & 0.0000005\% & 0.43\% & 0.0000003\%\\
RSAPKCS1 & 0.0000031\% & 0.001\% & 0.0000018\%\\
ECDSA & 0.0001365\% & 0.02\% & 0.0000467\%\\
\bottomrule
\end{tabular}
}
\label{tab:sign_verify_stats_excluding_high}
\end{table}

\begin{table}
\centering
\caption{Verification failure for valid signatures.}
\resizebox{0.55\linewidth}{!}{

\begin{tabular}{lll}
\toprule
Scheme & Fail Count & $Rate_{raw}$ \\
\midrule
RSAPKCS1 & 2488 & 0.000005\% \\
RSAPSS & 1980 &  0.000004\% \\
ECDSA & 7325 & 0.000003\% \\
\bottomrule
\end{tabular}
}
\label{tab:sign_sdc}
\end{table}

For encryption, we attributed the remaining 2\% of the RSA samples and all AES samples to transient/system errors, which have a occurrence rate of 0.00000003\%.
These samples appeared on a single device per failure of operation, suggesting that they could be due to defective hardware~\cite{hochschild2021cores}.
The majority of these samples (AES, RSA) that cannot be explained occur only on devices with limited memory: 53 low end, and 3 mid tier.
We used the known key to recalculate when available.
This revealed a total of three devices (2 AES \& 1 RSA) which do not compute the AES or RSA functions.
In one case, the encrypted message was successfully decrypted on the same device, even though it was generated with an invalid encryption function.

For HMAC, after filtering out the samples that are due to implementation bugs, we observed a 0.0000001\% chance of corrupted HMAC.

We identified a device that produced 13 invalid ECDH exchanges, likely due to hardware failure.
However, we did not identify any trend for ECDH, suggesting that newer devices that implement ECDH have lower failure rates.

\important{\textbf{Integrity (2):} Silent data corruption affects various cryptographic operations at a very low rate, but it occurs more often in lower-end devices.}

\section{Random Number Analysis} \label{sec:random}
We used the Paranoid library to analyze four types of random numbers that we extract from our collected cryptographic samples: 
ECDSA nonce, PSS salt, OAEP salt, and AES IV.
Previous work has shown that weakly generated parameters can lead to compromised keys~\cite{breitner2019biased}.

Paranoid implements several randomness tests~\cite{paranoidrandom} based on NIST SP 800-22~\cite{rukhin2001statistical} and some additional tests. 
The complete list is presented in Table~\ref{tab:paranoid_randomness}.
The test functions receive a large random number and output $p-value$(s), the lower a $p-value$, the weaker the random number source is.
Following the suggested default parameters in the library, we consider a $p-value < 1e-9$ as weak.

For each random parameter, we run two sets of tests.
One is based on grouping numbers based on unique device fingerprint, API level, and computation backing (TEE, Software, StrongBox) (sec~\ref{sec:random_fingerprint}).
Second, we group them by the unique device id hash and the computing backend. (sec~\ref{sec:random_device}).
Then we concatenate them to form a large sequence.

Our approach can identify weakly generated cryptographic random numbers, but we do not have visibility into how these parameters are generated.
Even if a device has a strong random number generator (RNG) (e.g., FIPS certified), software bugs or other misuse of the random entropy may result in weak parameters.

\begin{table}[t]
\centering
\caption{Supported randomness test by Paranoid.}
\resizebox{\linewidth}{!}{
\begin{tabular}{lp{4.7cm}l}
\toprule
Test Group & Description & Pvalues \\
\midrule
Frequency & NIST SP 800-22~\cite{rukhin2001statistical} (Sec 2.1) & 1 \\
BlockFrequency & NIST SP 800-22~\cite{rukhin2001statistical} (Sec 2.2) & 1 \\
Runs & NIST SP 800-22~\cite{rukhin2001statistical} (Sec 2.3) & 1 \\
LongestRuns & NIST SP 800-22~\cite{rukhin2001statistical} (Sec 2.4) & 1 \\
BinaryMatrixRank & NIST SP 800-22~\cite{rukhin2001statistical} (Sec 2.5) & 1 \\
Spectral & NIST SP 800-22~\cite{rukhin2001statistical} (Sec 2.6) & 1 \\
NonOverlappingTemplate & NIST SP 800-22~\cite{rukhin2001statistical} (Sec 2.7) & 588 \\
OverlappingTemplate & NIST SP 800-22~\cite{rukhin2001statistical} (Sec 2.8)& 1 \\
Universal & NIST SP 800-22~\cite{rukhin2001statistical} (Sec 2.9) & 1 \\
LinearComplexity & NIST SP 800-22~\cite{rukhin2001statistical} (Sec 2.10) & 8 \\
Serial & NIST SP 800-22~\cite{rukhin2001statistical} (Sec 2.11)& 40 \\
ApproximateEntropy & NIST SP 800-22~\cite{rukhin2001statistical} (Sec 2.12) & 14 \\
RandomWalk & NIST SP 800-22~\cite{rukhin2001statistical} (Sec 2.13-15) & 28 \\
FindBias & LLL-based approach~\cite{paranoidrandom}  & 5 \\
LargeBinaryMatrixRank & BinaryMatrixRank w/ large matrices~\cite{paranoidrandom}& 7 \\
LinearComplexityScatter & Linear complexity of interleaved bits~\cite{paranoidrandom} & 6 \\
\bottomrule
\end{tabular}

}
\label{tab:paranoid_randomness}
\end{table}

\subsection{Randomness Test Per Chipset Model} \label{sec:random_fingerprint}

\begin{table}[t]
\centering
\caption{Overview of testing random numbers across unique device fingerprints.}
\resizebox{\linewidth}{!}{
\begin{tabular}{llllll}
\toprule
Source & PValues & Sample Groups & Failed & Chipsets & Chipsets Failed \\
\midrule
PSS Salt    & 402M & 24,050 & 10 & 342 & 5\\
OAEP Salt   & 441M & 20,674 & 0 & 342 & 0\\
AES IV      & 348M & 20,335 & 2 & 335 & 2\\
ECDSA Nonce & 632M & 48,004 & 30 & 434 & 20\\
\bottomrule
\end{tabular}

}
\label{tab:random_per_fingerprint}
\end{table}

\begin{table*}[t]
\centering
\caption{Top chipsets that fail randomness test.}
\resizebox{\linewidth}{!}{

\begin{tabular}{llrrV{7cm}lll}
\toprule
Manufacturer & Chipset & Failed Test & Models & Test Groups & API Level & Random Source & Backing \\
\midrule
Mediatek & MT6739WW & 4252 & 1 & ApproximateEntropy, FindBias, LargeBinaryMatrixRank, LinearComplexity, LongestRuns, NonOverlappingTemplate, Serial, Spectral & 27 & ECDSA Nonce, PSS Salt & TEE \\
Mediatek & MT6580 & 85 & 6 & ApproximateEntropy, LargeBinaryMatrixRank, Serial & 24, 27 & AES IV, ECDSA Nonce, PSS Salt & Software, TEE \\
Mediatek & MT6580M & 76 & 6 & LargeBinaryMatrixRank, Serial & 27 & ECDSA Nonce, PSS Salt & TEE \\
Mediatek & MT6739CH & 48 & 1 & ApproximateEntropy, LargeBinaryMatrixRank, Serial & 27 & ECDSA Nonce & TEE \\
Qualcomm & SDM636 & 31 & 2 & ApproximateEntropy, Serial & 29 & ECDSA Nonce & Software \\
Qualcomm & MSM8996 & 28 & 1 & Serial & 29 & ECDSA Nonce & TEE \\
Qualcomm & SM6125 & 24 & 1 & ApproximateEntropy, Serial & 29 & ECDSA Nonce & Software \\
UNISOC & SC9832E & 22 & 1 & ApproximateEntropy, Serial & 30 & AES IV & TEE \\
Mediatek & MT6762D & 22 & 2 & Serial & 30 & ECDSA Nonce, PSS Salt & TEE \\
Rockchip & RK3566 & 21 & 1 & Serial & 30 & ECDSA Nonce & TEE \\
Qualcomm & SDM855 & 16 & 1 & ApproximateEntropy, Serial & 29 & ECDSA Nonce & Software \\
Qualcomm & QCM6490 & 8 & 1 & Serial & 32 & ECDSA Nonce & TEE \\
Rockchip & RK3399 & 8 & 1 & Serial & 28 & ECDSA Nonce & TEE \\
Mediatek & MT6797M & 6 & 1 & Serial & 23 & ECDSA Nonce & TEE \\
Mediatek & MT6765 & 6 & 1 & Serial & 31 & ECDSA Nonce & TEE \\
% Qualcomm & SDM845 & 6 & 1 & Serial & 30 & ECDSA Nonce & TEE \\
% Qualcomm & SDM660 & 4 & 2 & Serial & 30 & ECDSA Nonce & TEE \\
% Mediatek & MT6737T & 4 & 1 & Serial & 23 & ECDSA Nonce & Software \\
% Qualcomm & MSM8916 & 2 & 1 & Serial & 23 & ECDSA Nonce & Software \\
% Qualcomm & SDM439 & 2 & 1 & Serial & 29 & ECDSA Nonce & TEE \\
% Qualcomm & SDM710 & 2 & 1 & LargeBinaryMatrixRank & 30 & ECDSA Nonce & TEE \\
% Qualcomm & QCS4290 & 2 & 1 & Serial & 30 & PSS Salt & TEE \\
\bottomrule
\end{tabular}

}
\label{tab:random_top_failed}
\end{table*}

\begin{table*}[ht]
\centering
\caption{Top chipsets that fail randomness test evaluated per device.}
\resizebox{\linewidth}{!}{

\begin{tabular}{llrrrV{7cm}lll}
\toprule
Manufacturer & Chipset & Failed Test & Devices & Models & Test Groups & API Level & Random Source & Backing \\
\midrule
Mediatek & MT6739WW & 1593 & 1 & 1 & ApproximateEntropy, FindBias, LargeBinaryMatrixRank, LongestRuns, NonOverlappingTemplate, Serial, Spectral & 27 & ECDSA Nonce, PSS Salt & TEE \\
Samsung & s5e8825 & 87 & 2 & 1 & ApproximateEntropy, Serial, Spectral & 34 & ECDSA Nonce & TEE \\
Qualcomm & SDM660 & 62 & 4 & 2 & ApproximateEntropy, Serial, Spectral & 27, 28, 30 & ECDSA Nonce & TEE \\
Mediatek & MT6762D & 57 & 6 & 2 & ApproximateEntropy, LargeBinaryMatrixRank, Serial, Spectral & 30 & AES IV, ECDSA Nonce, OAEP Salt & TEE \\
Mediatek & MT6580 & 45 & 1 & 6 & LargeBinaryMatrixRank, Serial & 27 & ECDSA Nonce, PSS Salt & TEE \\
Mediatek & MT6580M & 34 & 1 & 1 & LargeBinaryMatrixRank, Serial & 27 & ECDSA Nonce & TEE \\
Mediatek & MT6765 & 30 & 3 & 1 & ApproximateEntropy, Serial, Spectral & 30, 31, 33 & ECDSA Nonce & TEE \\
Intel & i7-5500U & 28 & 1 & 6 & ApproximateEntropy, Serial, Spectral & 29 & OAEP Salt & Software \\
Mediatek & MT6833V/NZA & 26 & 1 & 1 & ApproximateEntropy, Serial, Spectral & 33 & ECDSA Nonce & TEE \\
Mediatek & MT6739CH & 24 & 1 & 1 & LargeBinaryMatrixRank, Serial & 27 & ECDSA Nonce & TEE \\
Samsung & Exynos 850 & 22 & 1 & 1 & ApproximateEntropy, Serial, Spectral & 33 & ECDSA Nonce & TEE \\
Mediatek & MT6781 & 20 & 1 & 1 & ApproximateEntropy, Serial, Spectral & 33 & ECDSA Nonce & TEE \\
Qualcomm & SDM450 & 20 & 2 & 2 & Serial, Spectral & 27, 28 & AES IV, ECDSA Nonce & TEE \\
Qualcomm & SM6115 & 18 & 1 & 1 & ApproximateEntropy, Serial, Spectral & 30 & ECDSA Nonce & TEE \\
Mediatek & MT6769 & 10 & 1 & 1 & Serial & 29 & ECDSA Nonce & TEE \\
Qualcomm & SM7325 & 6 & 1 & 1 & Serial & 33 & ECDSA Nonce & TEE \\
% Mediatek & MT6877V/TTZA & 4 & 1 & 1 & Serial & 34 & AES IV & TEE \\
% Qualcomm & SM6375 & 4 & 1 & 1 & Serial & 30 & ECDSA Nonce & TEE \\
% Mediatek & MT8788A & 2 & 1 & 1 & Serial & 29 & ECDSA Nonce & TEE \\
\bottomrule
\end{tabular}
}
\label{tab:random_top_failed_device}
\end{table*}

From each unique device model, we extracted random bytestrings of at least 640\,kB and applied a data processing pipeline that executes each test from  Table~\ref{tab:paranoid_randomness} outputting 704 p-values per 128\,kB block of random data.
This required running each test at least 5 times per unique device fingerprint to gain high confidence in our results.

Table~\ref{tab:random_per_fingerprint} gives an overview of the groups we have tested, resulting in hundreds of millions of p-values.
We consider a group of devices to have failed a test if the p-values are less than $1e-9$ for at least two attempts.
We observed tens of unique device fingerprints that have failed one or more randomness tests on different chipsets.

Table~\ref{tab:random_top_failed} highlights the chipsets with the highest number of failures from three manufacturers.
One of the main trends in these failures is related to Mediatek.
Four different Mediatek chipsets (MT6739WW, MT6580, MT6580M, MT6739CH) with 14 unique device fingerprints with API level 27 produced weak random numbers across three sources of random number.
The other noticeable trend is that several chipsets from Qualcomm with API level 29-30 have weak ECDSA nonces, affecting 8 unique device fingerprints across 6 chipsets.
These weak random numbers affect several generations of devices from these manufacturers.

\subsection{Randomness Test Per Device} \label{sec:random_device}
Next, we grouped the numbers per device.
This means that the test runs against numbers that are generated on the same device but not necessarily contiguously.

Since we have a limited number of samples per unique device, we tested samples from devices that generated at least 160\,kB of random data.
We use the same data pipeline as in the previous section to run tests for each 32\,kB block of random numbers to ensure that each test has been executed at least 5 times per device and random source.

Table~\ref{tab:random_top_failed_device} reports the chipsets with the highest number of failed tests.
Some of the same chipsets appeared in both tables, increasing the confidence that a number of unique devices and device models with these chipsets consistently generate poor random numbers.
We also observe that the highly nonrandom sample from Mediatek MT6739WW is from a single device, with a relatively very bad source of random numbers that has repeatedly failed many of the tests.

\subsection{A Poor Random Number Generator}
\begin{figure}[h]
     \centering
         \begin{subfigure}[b]{0.492\linewidth}
         \centering
         \includegraphics[width=\textwidth]{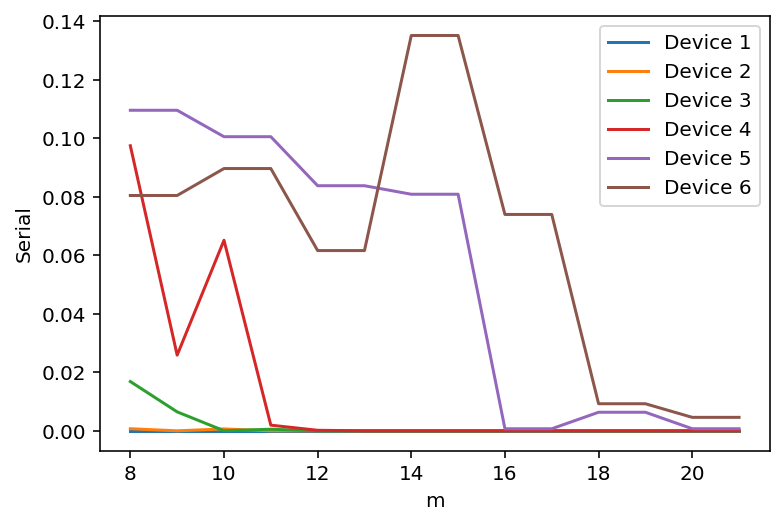}
         \caption{Trend of p-values from serial test per $m$.}
         \label{fig:serial_bad_random_m}
     \end{subfigure}
     \hfill
     \begin{subfigure}[b]{0.492\linewidth}
         \centering
         \includegraphics[width=\textwidth]{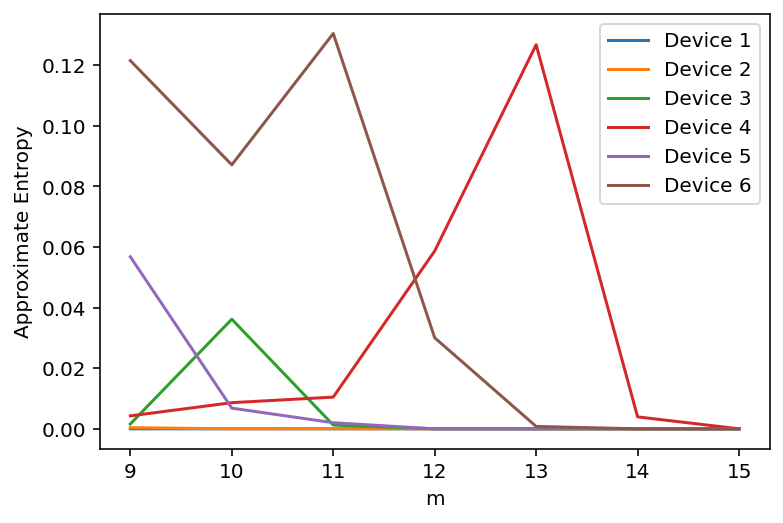}
         \caption{Trend of p-values from Approximate entropy test per $m$.}
         \label{fig:approximate_entropy_bad_random_m}
     \end{subfigure}
    \caption{Random number failures from Mediatek.}
\end{figure}

As a case study, we analyzed the p-values for  \texttt{Serial} and \texttt{ApproximateEntropy} tests.
These tests~\cite{rukhin2001statistical} output multiple p-values per parameter $m$, the window size.
We compared $p-values$ for increasing sizes of $m$ for various Mediatek devices in Table~\ref{fig:serial_bad_random_m} and \ref{fig:approximate_entropy_bad_random_m}.
For these bad random numbers, increasing values of $m$ result in a downward trend for the $p-value$ of these tests.

\begin{figure}[h]
     \centering
         \begin{subfigure}[b]{0.493\linewidth}
         \centering
         \includegraphics[width=\textwidth]{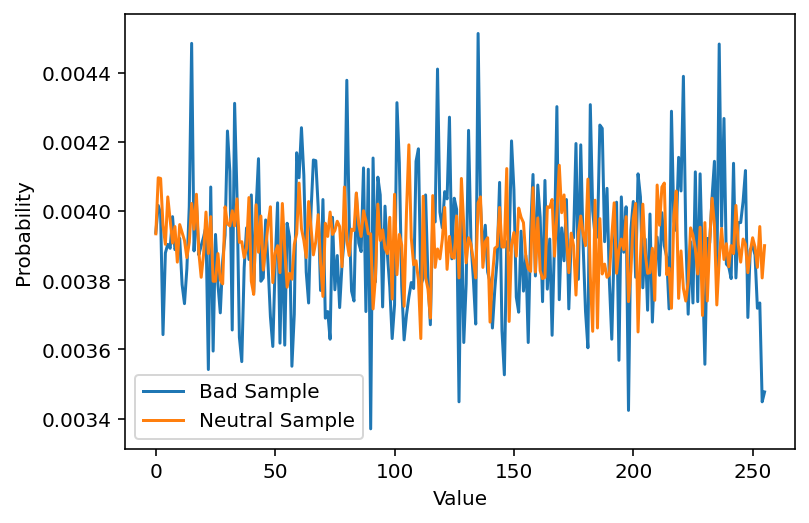}
         \caption{Probability of occurrence of different 8-bit values.}
         \label{fig:serial_bad_random_prob}
     \end{subfigure}
     \hfill
     \begin{subfigure}[b]{0.493\linewidth}
         \centering
         \includegraphics[width=\textwidth]{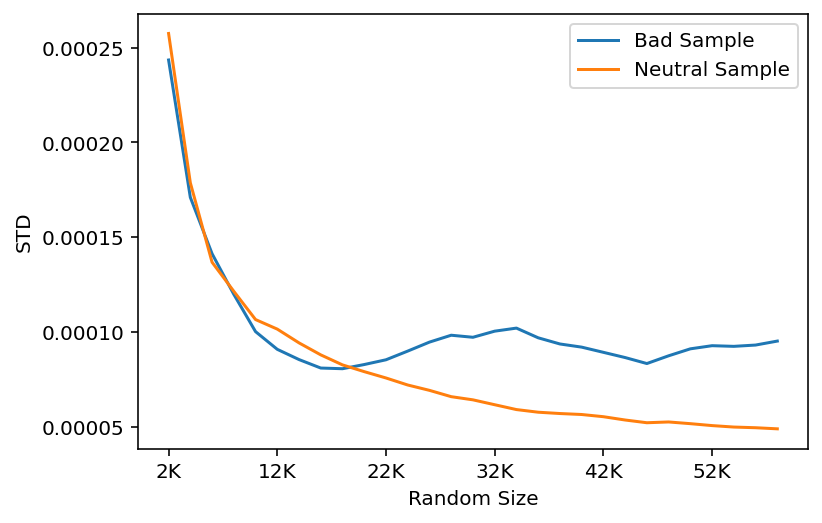}
         \caption{STD of the probability per sample size.}
         \label{fig:serial_bad_random_std}
     \end{subfigure}
    \caption{Random number fail serial test.}
\end{figure}

Next, we picked a 64\,kB sample of random bytes from Mediatek MT6739WW and calculated the probability of every possible 8-bit value.
Figure~\ref{fig:serial_bad_random_prob} compares these probability with a good RNG.
As we can see, for the weak source of random numbers, certain bytes have an unusually high or low probability ($blue$).
Looking at the standard deviation of the probabilities per sample size in Figure~\ref{fig:serial_bad_random_std}, we observe that, for these weak random numbers, the STD does not tend toward zero with more samples.
The weakness can be identified after about 12\,kB random bytes.

Our observations show that some devices from Mediatek consistently produce weak random numbers, weakening the security of cryptographic functions.

\important{
\textbf{Bug (11):} Weakly-generated random numbers for various cryptographic parameters affect several chipsets from at least two manufacturers with high confidence.
}

\section{Blind Cryptanalysis} \label{sec:paranoid}
In this section, we report our findings from analyzing public key artifacts and looking for vulnerabilities due to weakly generated keys~\cite{nemec2017return,bernstein2013factoring}.

\subsection{Public Key Cryptanalysis}
We executed 4 tests for EC and 16 tests for RSA keys against 526.2 million EC and 505 million RSA keys.
For EC keys, we did not identify any weakness, but for RSA, we identified 94 weak RSA keys.
Table~\ref{tab:paranoid_pk_tests} shows the number of weak keys identified for different tests.
When certain vulnerable bit patterns exist in the private key, the RSA parameters $p$ and $q$ can be factored from the public modulus.
The three tests that have identified the largest number of weak keys: \emph{BitPatterns}, \emph{PermutedBitPatterns}, and \emph{ContinuedFractions} are indicators of low-entropy RSA keys. 

\para{Corrupted RSA Key} We identified 8 keys that are from the Marvell BG4-CT TEE with API levels 26-28.
The recovered factors for these keys are not prime numbers, and have a low entropy and an invalid exponent.
The operations performed on these keys have failed, which indicates that these keys are generated by a buggy implementation that produces or stores corrupted RSA keys.

\para{StrongBox generates weak RSA keys}
The rest of the identified weak RSA keys are successfully factored by \emph{BitPatterns}, \emph{PermutedBitPatterns}, and \emph{ContinutedFractions}.
When we looked at the factors, the majority of the bits were zero.
We found that these keys were generated by devices with the Google Tensor chipset with an outdated firmware and API level 31.
We reported this vulnerability to Google Pixel and they confirmed that this vulnerability was already reported and fixed (CVE-2022-20117~\cite{gsacve}).

\begin{table}
\centering
\caption{Weak RSA keys identified by different tests.}
\resizebox{\linewidth}{!}{
\begin{tabular}{llllll}
\toprule
Check & Description & Weakness \\
\midrule
BitPatterns & Repeating bit pattern. & 94 \\
PermutedBitPatterns &  Repeating bit pattern similar to \cite{bernstein2013factoring}. & 94\\
ContinuedFractions & Large coefficients in a continued fraction~\cite{bernstein2013factoring}.& 86\\
Exponents & Exponent $e \neq 65537$. & 6 \\
ROCA & ROCA keys~\cite{nemec2017return}& 2 \\ 
ROCAVariant & Variant of ROCA keys~\cite{nemec2017return} & 0 \\
HighAndLowBitsEqual & $p$ and $q$ share sufficiently many high and low bits. & 0 \\ 
GCD & Batch GCD among several keys. & 0 \\ 
GCDN1 & $GCD(n_1 - 1, n_2 - 1) >= bound$~\cite{nemec2017return} & 0 \\
SmallUpperDifferences & $abs(p - q)$ has some special form. & 0 \\
KeypairDenylist & CVE-2021-41117 vulnerability~\cite{GHSL} & 0 \\
LowHammingWeight & $p$ and $q$ have a low Hamming weight. & 0 \\
Pollardpm1 & Number preceding the factor, $p-1$, is power-smooth~\cite{pollard1974theorems}. & 0 \\
UnseededRand & Keys were generated with unseeded PRNGs~\cite{GHSL} & 0 \\
Fermat & Fermat's factorization method~\cite{lehman1974factoring}& 0 \\
OpensslDenylist & Debian OpenSSL Predictable PRNG (CVE-2008-0166) & 0 \\
\bottomrule
\end{tabular}

% Fermat's factorization method is based on the representation of an odd
% integer as the difference of two squares:
%   n = a**2 - b**2.
% That difference is algebraically factorable as n = pq = (a + b)(a - b).
% See https://en.wikipedia.org/wiki/Fermat%27s_factorization_method

%   This test computes the continued fraction of each RSA modulus with a power of two. 
  
%   The test fails if an unusually large coefficient is found.
%   "Factoring RSA keys from certified smart cards: Coppersmith in the wild"
%   https://smartfacts.cr.yp.to/smartfacts-20130916.pdf
%   Many of the weak primes in Appendix A have a repetitive pattern.
%   Such primes have the property that they are close to a/b * 2^m,
%   where a/b is a small fraction and m is the size of the prime.
%   If the both prime factors have this form then their product n
%   also has this form if the fractions are small enough.
%   Computing a continued fraction can detect such forms: a modulus n
%   is close to some a/b * 2^m if one of the coefficients of the continued
%   fraction is unusually large.
  
% ECKeySmallDifference
% WeakCurve
% ValidECKey
% WeakECPrivateKey

% PermutedBitPatterns 94
% BitPatterns 94

% Tries to 

}
\label{tab:paranoid_pk_tests}
\end{table}

\important{
\textbf{Bug (12):} Corrupted RSA public keys affect one chipset.

\noindent\textbf{Bug (13):} Vulnerable RSA keys due to weak prime factors affect one StrongBox implementation.
}

\section{ECDSA Timing Side channel} \label{sec:ecdsatiming}
In this section, we analyze ECDSA signatures for a class of side-channel attacks, where the correlation between ECDSA timing and the size of the nonce can be exploited using lattice-based cryptanalysis techniques~\cite{moghimi2020tpm,jancar2020minerva,brumley2011remote}.

\subsection{Modeling Timing Side Channel}
We focus on the case where a short ECDSA nonce---most significant bits are zero---results in faster execution~\cite{jancar2020minerva,moghimi2020tpm}.
For simplicity, we simulate the leakage for the case where for every $1/256 = 0.0039$ of the signatures, one signature on average has a nonce that is less than $32$ bytes long.
Since most of the execution time for ECDSA is spent on scalar multiplication with the nonce~\cite{li2013novel}, we hypothesize that a nonconstant-time implementation that completes on average for $m$ milliseconds will be at least $1/32 = 0.03125$ faster when the ECDSA nonce is one or more bytes short.
Based on this hypothesis, we simulate the ideal correlation between the ECDSA nonce and noisy timing measurement over a standard Gaussian distribution.

The average execution time for ECDSA for various devices is less than 500 ms, and most implementations have an average of less than 100 ms (Figure~\ref{fig:ecdsa_wall_time_avg} in the appendix).
For example, consider the distribution for StrongBox in one device with an average signature time of 87 ms and the standard deviation of 10 ms (Figure~\ref{fig:ecdsa_sample_timing} in the Appendix).
The large standard deviation (STD) compared to the average implies a higher variability of execution times and potentially more noise in the execution times.

\begin{figure}
     \centering
         \begin{subfigure}[b]{0.468\linewidth}
         \centering
         \includegraphics[width=\textwidth]{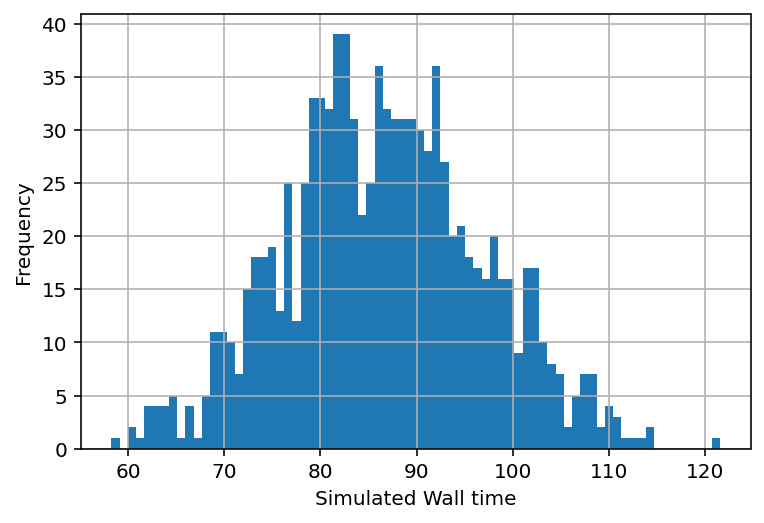}
         \caption{Simulated timing over a Gaussian distribution.}
         \label{fig:ecdsa_timing_simulated}
     \end{subfigure}
     \hfill
     \begin{subfigure}[b]{0.510\linewidth}
         \centering
         \includegraphics[width=\textwidth]{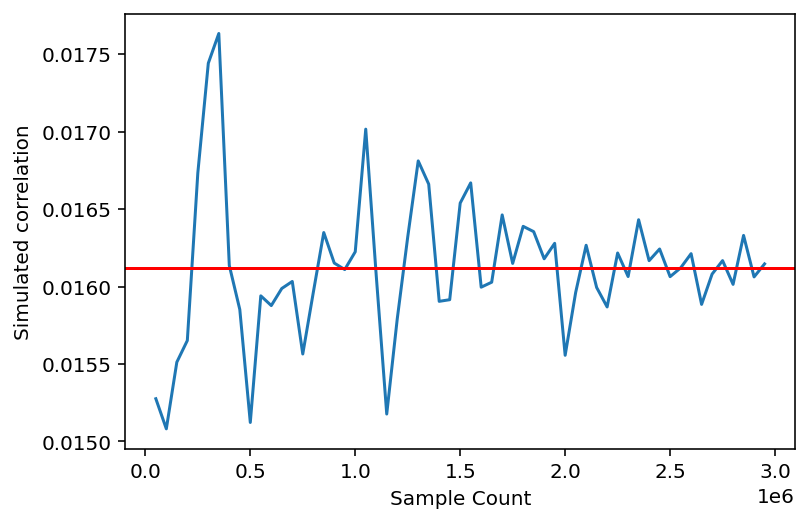}
         \caption{Correlation coefficient for the simulated timing leakage.}
         \label{fig:ecdsa_timing_corr_simulated}
     \end{subfigure}
    \caption{Simulating timing leakage.}
\end{figure}

We simulate the timing-leakage correlation as follows: 
First, based on the average and STD of the timing for a device, we calculate random timing samples over the Gaussian distribution, as in Figure~\ref{fig:ecdsa_timing_simulated}.
Then, we inject the timing delay by subtracting the timing for $0.0039$ of the timing samples with $1/32$ of the average execution time.
Figure~\ref{fig:ecdsa_timing_corr_simulated} shows the correlation coefficient (\texttt{corr}) for the Gaussian distribution mentioned earlier (average: 87 ms, std: 10 ms).
By increasing the number of samples, \texttt{corr} converges to a target, in this case $0.016$.
We later use this model to identify whether a device has a timing behavior that resembles the leakage.
If the \texttt{corr} of the execution timing and the byte length are close to or more than the simulated \texttt{corr}, it is likely that the device is vulnerable to ECDSA timing attacks.

\subsection{Real-world Evaluation}
\FrameWorkName\ applies this model to ECDSA samples by grouping them by device, since the timing may vary a lot on different devices, due to different software and power profiles.
We take samples from devices that have successfully computed and verified at least 10,000 ECDSA signatures with the fixed EC key.
For these samples, we extract the ECDSA nonce, similar to ~\ref{sec:random_fingerprint}.
We then calculate the \texttt{corr} between the byte length of the ECDSA nonce and the execution time.

We identified more than 3,000 devices that match the timing leakage model we simulated earlier.
Heuristically, we picked samples that have $\texttt{corr} > 0.005$, which accounted for $4818$ of the $131,755$ samples tested.
We further filter those that are $\texttt{corr} > \texttt{simulated\_corr} - 0.001$, and left with $3159$ samples.
Table~\ref{tab:ecdsa_timing_top_devices} summarizes the chipsets with the highest number of potentially vulnerable devices.
Qualcomm has several chipsets across various API levels that appear in the top list and are potentially vulnerable to ECDSA timing side channels.

\begin{table}
\centering
\caption{Chipsets with the highest number of failing ECDSA timing side-channel test.}
\resizebox{\linewidth}{!}{
% \begin{tabular}{llllrrl}
% \toprule
% Make & Model & API Level & Backing & Weak & Total & $Rate_{device}$ \\
% \midrule
% Manufacturer B & Chipset 15 & 25, 27, 28, 29, 30, 32 & TEE & 1128 & 16007 & 1.16\% \\
% Manufacturer B & Chipset 32 & 28, 29 & TEE & 245 & 2964 & 0.25\% \\
% Manufacturer B & Chipset 31 & 29, 30 & TEE & 237 & 2297 & 0.24\% \\
% Manufacturer B & Chipset 30 & 30, 31, 32, 33, 34 & TEE & 201 & 1677 & 0.21\% \\
% Manufacturer A & Chipset 7 & 30 & TEE & 106 & 483 & 0.11\% \\
% Manufacturer G & Chipset 39 & 28, 30 & TEE & 98 & 469 & 0.10\% \\
% Manufacturer B & Chipset 35 & 30, 31, 33, 34 & TEE & 78 & 1086 & 0.08\% \\
% Manufacturer B & Chipset 40 & 28, 29, 30 & TEE & 73 & 2408 & 0.07\% \\
% Manufacturer B & Chipset 41 & 30 & TEE & 68 & 638 & 0.07\% \\
% Manufacturer A & Chipset 42 & 28, 29, 30, 31, 33 & TEE & 56 & 323 & 0.06\% \\
% Manufacturer B & Chipset 20 & 27, 28 & TEE & 52 & 280 & 0.05\% \\
% Manufacturer E & Chipset 43 & 28 & TEE & 48 & 371 & 0.05\% \\
% Manufacturer B & Chipset 38 & 34 & Strongbox & 44 & 5279 & 0.05\% \\
% Manufacturer A & Chipset 44 & 27, 28, 29 & TEE & 42 & 263 & 0.04\% \\
% Manufacturer A & Chipset 45 & 28, 29, 31 & TEE & 41 & 340 & 0.04\% \\
% Manufacturer B & Chipset 37 & 26, 28, 29 & TEE & 39 & 916 & 0.04\% \\
% Manufacturer B & Chipset 46 & 31, 33, 34 & Strongbox & 39 & 6414 & 0.04\% \\
% Manufacturer B & Chipset 21 & 26, 28, 29, 30, 33 & Software, TEE & 36 & 1089 & 0.04\% \\
% Manufacturer B & Chipset 36 & 28, 29 & TEE & 33 & 3804 & 0.03\% \\
% Manufacturer B & Chipset 14 & 28, 29 & TEE & 32 & 305 & 0.03\% \\
% \bottomrule
% \end{tabular}

\begin{tabular}{llllrrl}
\toprule
Make & Model & API Level & Backing & Weak & Total & $Rate_{device}$ \\
\midrule
Qualcomm & SDM660 & 25, 27, 28, 29, 30, 32 & TEE & 1128 & 16007 & 1.16\% \\
Qualcomm & SDM450 & 28, 29 & TEE & 245 & 2964 & 0.25\% \\
Qualcomm & SM6115 & 29, 30 & TEE & 237 & 2297 & 0.24\% \\
Qualcomm & SM6375 & 30, 31, 32, 33, 34 & TEE & 201 & 1677 & 0.21\% \\
Mediatek & MT6762D & 30 & TEE & 106 & 483 & 0.11\% \\
MStar & T22 & 28, 30 & TEE & 98 & 469 & 0.10\% \\
Qualcomm & SM4350 & 30, 31, 33, 34 & TEE & 78 & 1086 & 0.08\% \\
Qualcomm & SDA660 & 28, 29, 30 & TEE & 73 & 2408 & 0.07\% \\
Qualcomm & QCM4290 & 30 & TEE & 68 & 638 & 0.07\% \\
Mediatek & MT6762 & 28, 29, 30, 31, 33 & TEE & 56 & 323 & 0.06\% \\
Qualcomm & SDM636 & 27, 28 & TEE & 52 & 280 & 0.05\% \\
Samsung & Exynos 8895 & 28 & TEE & 48 & 371 & 0.05\% \\
Qualcomm & SM8650 & 34 & Strongbox & 44 & 5279 & 0.05\% \\
Mediatek & MT6763 & 27, 28, 29 & TEE & 42 & 263 & 0.04\% \\
Mediatek & MT6761 & 28, 29, 31 & TEE & 41 & 340 & 0.04\% \\
Qualcomm & MSM8998 & 26, 28, 29 & TEE & 39 & 916 & 0.04\% \\
Qualcomm & SM8450 & 31, 33, 34 & Strongbox & 39 & 6414 & 0.04\% \\
% Qualcomm & MSM8996 & 26, 28, 29, 30, 33 & Software, TEE & 36 & 1089 & 0.04\% \\
% Qualcomm & MSM8953 & 28, 29 & TEE & 33 & 3804 & 0.03\% \\
% Qualcomm & SDM845 & 28, 29 & TEE & 32 & 305 & 0.03\% \\
\bottomrule
\end{tabular}

}
\label{tab:ecdsa_timing_top_devices}
\end{table}

\important{\textbf{Bug (14):} ECDSA timing side channels affects several chipsets from at least one manufacturer.}

\para{Discussion}
Although our methodology and this conservative threshold, motivated by end-to-end (e2e) exploits, such as Minerva~\cite{jancar2020minerva} and TPM-Fail~\cite{moghimi2020tpm}, effectively isolate the most critical threats, it may not detect all vulnerable implementations; catching every edge case would require more precise lab-controlled testing. 
%conc
Consequently, our findings serve as a valuable initial screen, which can now guide more targeted, in-depth follow-up analysis.

\subsection{ECDSA Timing Leak Case Study}
\begin{figure}
     \centering
     \includegraphics[width=0.6\linewidth]{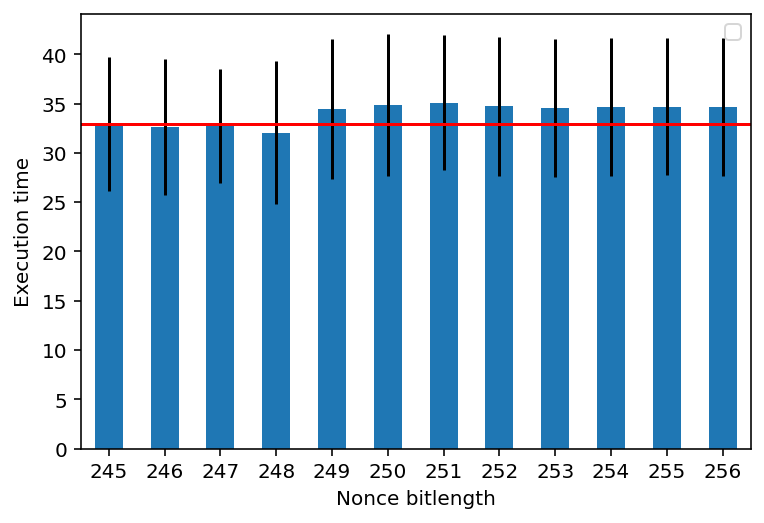}
     \caption{Execution time over nonce bit lengths.}
     \label{fig:ecdsa_timing_sample_errorbar}
\end{figure}

We analyzed one of the affected \emph{SDM660} devices.
Figure~\ref{fig:ecdsa_timing_sample_errorbar} shows the average execution time for a varying number of nonce lengths, highlighting that there is a 2 ms time difference as soon as the ECDSA nonce is short of 8 bits.

\begin{figure}
     \centering
         \begin{subfigure}[b]{0.493\linewidth}
         \centering
         \includegraphics[width=\textwidth]{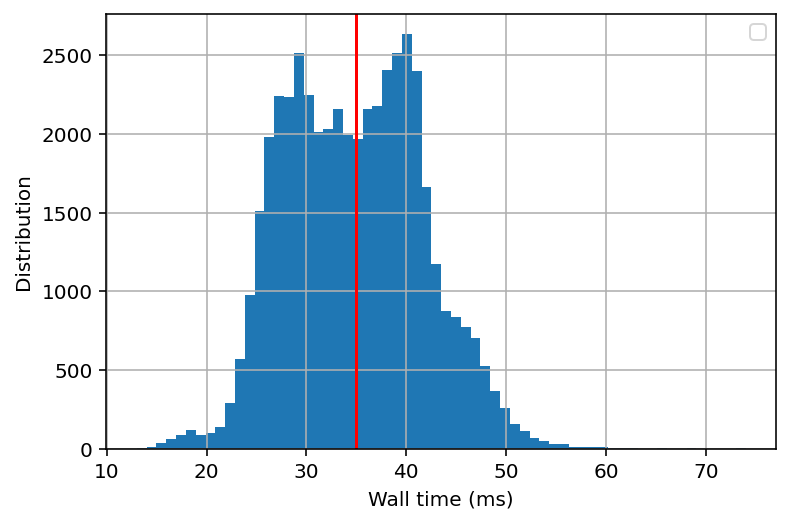}
         \caption{Timing distribution of ECDSA signing.}
         \label{fig:ecdsa_timing_sample_bad}
     \end{subfigure}
     \hfill
     \begin{subfigure}[b]{0.493\linewidth}
         \centering
         \includegraphics[width=\textwidth]{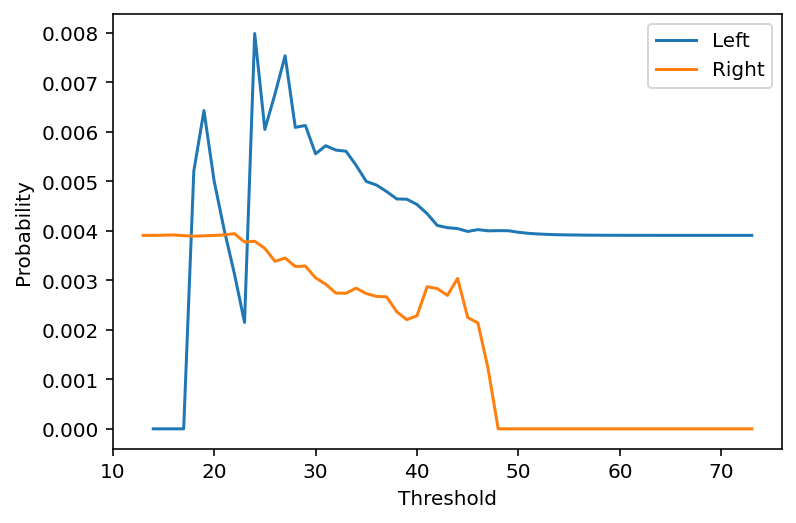}
         \caption{Probability of short nonce per splitting the samples.}
         \label{fig:ecdsa_timing_sample_bad_prob}
     \end{subfigure}
    \caption{Vulnerable ECDSA timing.}
\end{figure}

Based on this device's execution time distribution (Figure \ref{fig:ecdsa_timing_sample_bad}), we calculated the probability of samples that are less than 32 bytes with different timing thresholds.
As we can see in Figure~\ref{fig:ecdsa_timing_sample_bad_prob}, even though we expect this to be probably close to $1/256 = 0.0039$, it deviates much from it from 25 to 45 ms, implying that the timing depends on the byte length of the ECDSA nonce.
 
\begin{figure}[t]
     \centering
         \begin{subfigure}[b]{0.493\linewidth}
         \centering
         \includegraphics[width=\textwidth]{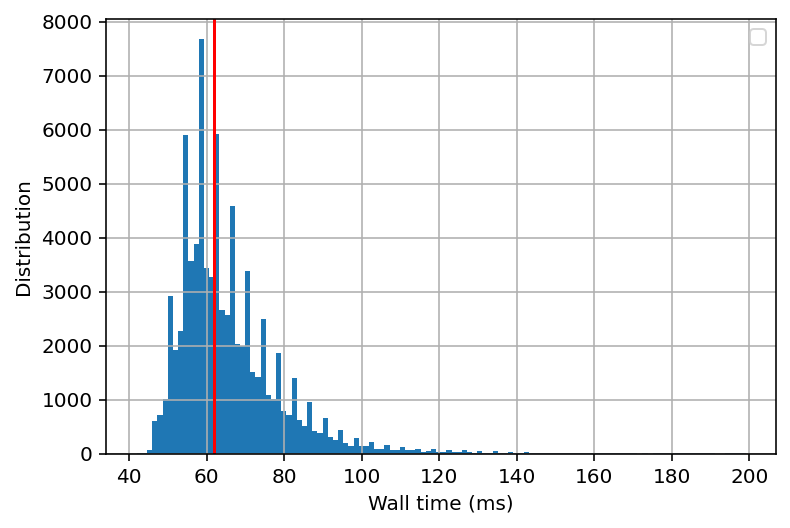}
         \caption{Timing distribution of ECDSA signing.}
         \label{fig:ecdsa_timing_sample_good}
     \end{subfigure}
     \hfill
     \begin{subfigure}[b]{0.493\linewidth}
         \centering
         \includegraphics[width=\textwidth]{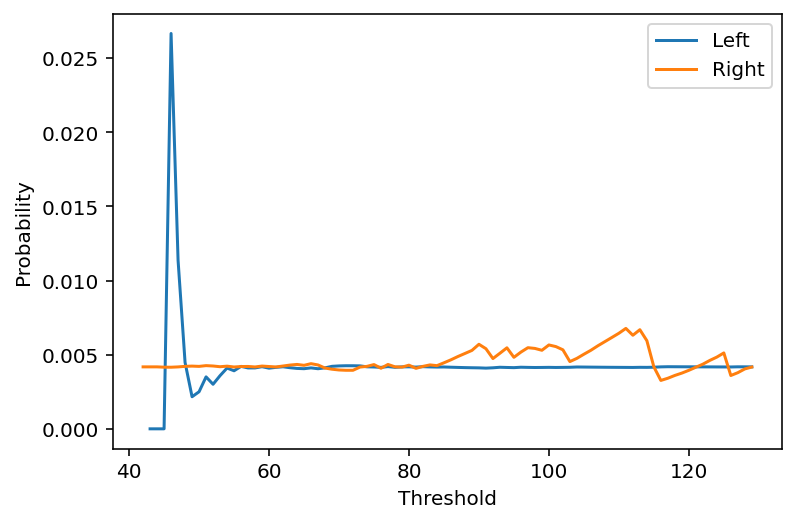}
         \caption{Probability of short nonce per splitting the samples.}
         \label{fig:ecdsa_timing_sample_good_prob}
     \end{subfigure}
    \caption{Neutral ECDSA timing.}
\end{figure}

In contrast, if we calculate the probability for a neutral sample from a a constant-time implementation (Figure \ref{fig:ecdsa_timing_sample_good} and \ref{fig:ecdsa_timing_sample_good_prob}), the probability is consistent and approximate around $0.0039$ disregarding the chosen threshold.

\section{Lessons learned \& Countermeasures} \label{sec:discuss}
\para{Hardware performance}
Hardware-backed cryptography (StrongBox) with protection against physical attacks is on the rise, but its adoption has performance implications (Section \ref{sec:perf}).
The widening performance gap compared to software may prohibit their adoption for latency-critical apps.
For example, the average execution time for RSA signing is of the order of $4-5 \times$ slower than TEE.
Countermeasures against physical attacks may impose more performance overheads for StrongBox.
Improving the performance of hardware implementations is necessary, especially with the introduction of PQC schemes with more demanding compute and memory requirements~\cite{ducas2018crystals}.

\para{Compliance test}
A large number of bugs that affect the availability of cryptographic functions are due to the incompatibility of the implementation with the Android API (\emph{Bug 1, 4, 5}).
This has surfaced in forms of failure to support the correct key format and functions despite API requirements.
These errors would make it difficult for developers to deploy apps that benefit from the Android keystore.
They have to rely on less secure alternatives, e.g. pure software, which do not have protection.
We suggest a rigorous compliance testing that covers cryptographic functions to improve the ecosystem around such bugs~\cite{androidcts}.

\para{Memory safety}
Software bugs due to memory errors have resulted in unusable cryptographic functions due to key storage issues or misbehaving artifacts (Sections~\ref{sec:errors} and \ref{sec:functional} -- \emph{Bug 2, 3, 6-10}).
Cases such as generating signatures with the wrong key or contiguous blocks of zeros are potentially due to memory bugs.
Although these bugs may appear benign, they can expose cryptographic materials and circumvent the expected protection provided by the Android keystore.
Memory safe languages~\cite{rustlang} and verified cryptographic implementations~\cite{zinzindohoue2017hacl} can prevent these bugs.

\para{Fault tolerance}
The rare transient errors potentially contribute to the computation of a number of bad cryptographic artifacts.
Transient errors may degrade the security of cryptographic schemes~\cite{sullivan2022open}.
In particular, we observed that a number of invalid signatures were verified on the device, suggesting that some software implementations are not fault-tolerant.
Although fault resistance has traditionally been applied to protect against physical attacks, it is also important for software implementations as transient faults may grow with increasing density of transistors.

\para{Transparency}
We applied several cryptanalysis techniques (Section~\ref{sec:random} \& \ref{sec:paranoid}).
Despite doing this in a noisy environment where device behavior may vary, our analysis revealed a confirmed vulnerability and also identified several weak sources of random numbers.
In a controlled environment, an attacker can increase their chance by exercising these cryptographic functions with greater frequency and precision~\cite{bernstein2013factoring,nemec2017return}.
These vulnerabilities show that closed-source and proprietary cryptography is detrimental to security.

Several devices with the same hardware profile and the latest API level expose timing behavior and weaken ECDSA private keys.
Although this class of critical vulnerabilities has been known~\cite{brumley2011remote,moghimi2020tpm,jancar2020minerva}, it is very difficult to mitigate them.
We leave further analysis of them in a lab setting for future work, but such flaws can go unnoticed in closed-source software for years.

\para{Patching}
Some of the issues we identified are only present in older Android APIs or firmware, highlighting that deploying patches are not uniform across different devices.
At the time this publication, Google has officially ended security support for Android below API level 33, which means means devices running older Android are no longer protected by Google against newly discovered vulnerabilities.
While device manufacturers (OEMs) could backport security patches to their devices, this is not guaranteed and often does not happen for older models. 

\section{Conclusion} \label{sec:conclusion}

DroidCCT demonstrates that large-scale telemetry can serve as an effective, proactive defense, transforming billions of field samples into actionable security intelligence to guide targeted remediation and significantly raise the security baseline for billions of users.

\section*{Acknowledgements}
We would like to thank Luca Invernizzi, Shawn Willden, Paul Crowley, David Kleidermacher, and Amanda Walker from Google for their invaluable feedback and ongoing support.
We also acknowledge Daniel Bleichenbacher for his prior contribution to the development of the Paranoid~\cite{bbparanoid}, which was instrumental to the success of this study.

\bibliographystyle{IEEEtran}
\bibliography{references}

\newpage

\appendix
\subsection{Additional Figures}

\begin{table*}[t]
  \centering
\caption{Android Keystore features and their availability across different API levels.}
\label{tab:keystore_features}
\resizebox{.78\linewidth}{!}{
\begin{tabular}{lll}
\toprule
\textbf{Feature} & \textbf{API Level} & \textbf{Description} \\
\midrule
KeyInfo.isInsideSecureHardware & $<$ 31 & Returns true if the key is stored in Strongbox or TEE. \\ 
KeyInfo.getSecurityLevel & $>=$ 31 & Returns Software (0), TEE (1), StrongBox (2) enumerations. \\
KeyProperties.PURPOSE\_AGREE\_KEY & $>=$ 31 & Supports key exchange, only applicable to the ECDH scheme.\\
KeyProtection\$Builder & $>=$ 23 & Key definition for an importing keys. \\
KeyProtection\$Builder.setIsStrongBoxBacked & $>=$ 31 & Enable strongbox for imported keys.\\ 
KeyGenParameterSpec\$Builder & $>=$ 23 & Key definition for generating keys.\\ 
KeyGenParameterSpec\$Builder.setIsStrongBoxBacked & $>=$ 28 & Enable strongbox for key generation. \\ 
FEATURE\_STRONGBOX\_KEYSTORE & $>=$ 28 & Returns true if device supports Strongbox. \\ 
\bottomrule
\end{tabular}

}
\end{table*}

\begin{table*}[t]
\centering
\caption{Error trends across various cryptographic operations.}
\resizebox{\linewidth}{!}{

\begin{tabular}{llllllllllllll}
\toprule
Error & Operation                 & Key used        & Exception & Error message           & API Level & Computation Backing & Make & Model Count & Device Count & Min & Max & Median \\
\midrule
1 & ECDSA Sign                & Import/Generate & KeyPermanentlyInvalidatedException & Key permanently invalidated                              & 31 & TEE & 2 & 3 & 157 & 0.13\% & 0.26\% & 0.24\% \\
2 & ECDSA Sign                & Import/Generate & KeyStoreConnectException           & Failed to communicate with keystore service              & 32 & TEE & 1 & 1 & 97 & 0.61\% & 0.99\% & 0.80\% \\
3 & ECDSA Sign                & Generate        & NullPointerException               & method android.os.IBinder IKeystoreService.asBinder()    & 29 & TEE & 2 & 2 & 42 & 8.54\% & 13.92\% & 11.23\% \\
4 & RSA/ECDSA Sign            & Import/Generate & InvalidKeyException                & Keystore operation failed (Code 22)                      & 34 & TEE & 1 & 1 & 4,200 & 0.41\% & 19.12\% & 2.83\% \\
5 & ECDSA Sign                & Import/Generate & InvalidKeyException                & Keystore operation failed (Code -32)                     & 29 & TEE & 1 & 1 & 59 & 0.21\% & 0.23\% & 0.22\% \\
6 & RSA/ECDSA Sign            & Import/Generate & InvalidKeyException                & Keystore operation failed (Code -49)                     & 27 & TEE & 2 & 2 & 404 & 0.06\% & 0.15\% & 0.07\% \\
7 & RSA/ECDSA Sign            & Import/Generate & InvalidKeyException                & Keystore operation failed (Code -62)                     & 29 & TEE & 3 & 6 & 2,300 & 0.03\% & 94.17\% & 49.67\% \\
8 & RSA/ECDSA Sign            & Import/Generate & InvalidKeyException                & Keystore operation failed (Code -64)                     & 29, 30 & TEE & 1 & 2 & 71 & 0.06\% & 1.54\% & 1.19\% \\
9 & RSA Sign (PSS)             & Import/Generate & InvalidKeyException                & Keystore operation failed (Code -79)                     & 34 & Strongbox & 1 & 1 & 27 & 0.06\% & 0.10\% & 0.08\% \\
\textbf{10} & \textbf{RSA/ECDSA Sign}            & \textbf{Import/Generate} & \textbf{InvalidKeyException}                & \textbf{Invalid key blob}                                         & \textbf{28, 29} & \textbf{TEE} & \textbf{4} & \textbf{20} & \textbf{1,300} & \textbf{0.87\%} & \textbf{100.00\%} & \textbf{83.77\%} \\
11 & ECDSA Sign                & Import/Generate & InvalidKeyException                & Invalid key blob                                         & 33 & TEE & 1 &  1 & 38 & 0.15\% & 0.20\% & 0.17\% \\
12 & RSA Sign (PKCS1)           & Generate        & InvalidKeyException                & Memory allocation failed                                 & 33 & TEE & 1 &  1 & 11 & 0.69\% & 0.69\% & 0.69\% \\
13 & ECDSA Sign                & Import/Generate & InvalidKeyException                & Signature/MAC verification failed                        & 28 & TEE & 1 &  1 & 115 & 6.80\% & 20.00\% & 19.44\% \\
14 & RSA/ECDSA Sign            & Import/Generate & InvalidKeyException                & System error                                             & 34 & Strongbox & 1 &  1 & 157,300 & 0.99\% & 7.39\% & 2.95\% \\
15 & ECDSA Sign                & Generate        & InvalidKeyException                & Too many operations                                      & 30 & TEE & 1 &  2 & 55 & 1.23\% & 17.85\% & 2.00\% \\
16 & RSA/ECDSA Sign            & Import/Generate & InvalidKeyException                & Unknown error                                            & 26, 29, 30 & TEE & 3 & 4 & 1,800 & 0.04\% & 14.87\% & 0.09\% \\
17 & RSA Sign (PSS)             & Generate        & SignatureException                 & Code -1                                                  & 24 & TEE & 1 &  1 & 15 & 0.93\% & 0.93\% & 0.93\% \\
18 & RSA Sign (PSS)/ECDSA       & Import/Generate & SignatureException                 & Code -49                                                 & 27 & Software & 1 &  1 & 10,200 & 0.14\% & 3.55\% & 2.59\% \\
19 & RSA/ECDSA Sign            & Import/Generate & SignatureException                 & Invalid input length                                     & 25-31 & TEE & 4 & 21 & 2,000 & 0.07\% & 58.82\% & 8.02\% \\
20 & RSA Sign (PKCS1, PSS)      & Import/Generate & SignatureException                 & Invalid input length                                     & 33 & TEE & 1 &  2 & 1,200 & 0.73\% & 29.30\% & 7.91\% \\
21 & RSA/ECDSA Sign            & Import/Generate & SignatureException                 & Memory allocation failed                                 & 33 & TEE & 1 &  1 & 82 & 1.57\% & 5.37\% & 2.04\% \\
22 & RSA/ECDSA Sign            & Generate        & SignatureException                 & Unknown error                                            & 26, 29-32 & Strongbox/TEE & 2 & 6 & 433 & 0.11\% & 11.19\% & 1.37\% \\
23 & RSA Sign (PKCS1, PSS)      & Import/Generate & SignatureException                 & Unknown error                                            & 33 & Strongbox & 2 & 2 & 902 & 1.00\% & 4.10\% & 2.18\% \\
24 & RSA Encrypt (OAEP)        & Generate        & IllegalBlockSizeException          & input must be under 0 bytes                              & 26 & TEE & 1 & 1 & 14 & 0.09\% & 0.09\% & 0.09\% \\
25 & RSA Decrypt (PKCS1)       & Generate        & BadPaddingException                & Invalid argument                                         & 28, 29, 31 & TEE & 1 &  2 & 55 & 2.11\% & 3.74\% & 2.40\% \\
26 & RSA Decrypt (PKCS1)       & Generate        & BadPaddingException                & Invalid argument                                         & 33 & TEE & 1 &  1 & 16 & 2.87\% & 2.87\% & 2.87\% \\
27 & RSA Decrypt (PKCS1, OAEP) & Import/Generate & IllegalBlockSizeException          & Code -1                                                  & 24, 26, 29 & TEE & 2 & 3 & 248 & 0.05\% & 6.27\% & 0.19\% \\
\textbf{28} & \textbf{RSA Decrypt (PKCS1)}       & \textbf{Generate}        & \textbf{IllegalBlockSizeException}          & \textbf{Code -46}                                                 & \textbf{33} & \textbf{Strongbox} & \textbf{1} &  \textbf{1} & \textbf{294} & \textbf{39.02\%} & \textbf{58.33\%} & \textbf{52.66\%} \\
29 & RSA Decrypt (PKCS1, OAEP) & Import          & IllegalBlockSizeException          & Code -65535                                              & 23 & Software & 1 & 1 & 55 & 1.41\% & 2.03\% & 1.62\% \\
30 & RSA Decrypt (PKCS1, OAEP) & Import/Generate & IllegalBlockSizeException          & Invalid operation handle                                 & 25, 30 & TEE & 2 & 2 & 231 & 0.58\% & 1.46\% & 0.74\% \\
31 & RSA Decrypt (PKCS1, OAEP) & Import          & IllegalBlockSizeException          & Memory allocation failed                                 & 27 & TEE & 1 &  1 & 37 & 0.06\% & 0.08\% & 0.07\% \\
32 & RSA Decrypt (PKCS1, OAEP) & Import/Generate & IllegalBlockSizeException          & Memory allocation failed                                 & 33 & TEE & 1 &  1 & 277 & 1.34\% & 2.69\% & 1.83\% \\
33 & RSA Decrypt (PKCS1, OAEP) & Import/Generate & IllegalBlockSizeException          & Unknown error  & 26-32 & Strongbox/Software/TEE & 6 & 25 & 3,300 & 0.04\% & 42.86\% & 4.71\% \\
34 & RSA Decrypt (PKCS1, OAEP) & Import/Generate & IllegalBlockSizeException          & Unknown error  & 33 & Strongbox/TEE & 4 & 8 & 6,400 & 0.60\% & 37.50\% & 2.96\% \\
35 & RSA Decrypt (PKCS1, OAEP) & Import/Generate & IllegalBlockSizeException          & KeyStoreException: Unsupported purpose & 33 & Strongbox & 1 &  1 & 603 & 1.75\% & 4.67\% & 2.92\% \\
36 & AES Encrypt               & Import          & InvalidKeyException                & Keystore operation failed & 27 & Software & 1 &  1 & 445 & 1.86\% & 2.66\% & 2.26\% \\
37 & AES Encrypt               & Import          & InvalidKeyException                & Key not found        & 26 & Software & 1 &  1 & 14 & 9.81\% & 9.81\% & 9.81\% \\
38 & AES Encrypt               & Import/Generate & InvalidKeyException                & System error         & 34 & Strongbox & 1 &  1 & 13,000 & 3.67\% & 17.49\% & 8.60\% \\
39 & AES Decrypt               & Import/Generate & KeyStoreConnectException           & Failed to communicate with keystore service  & 29 & TEE & 2 & 2 & 35 & 1.52\% & 1.58\% & 1.55\% \\
40 & AES Decrypt               & Import/Generate & InvalidKeyException                & Keystore operation failed & 34 & TEE & 1 &  1 & 44 & 1.93\% & 2.13\% & 2.03\% \\
41 & AES Decrypt               & Import          & InvalidKeyException                & Key not found        & 26 & Software & 1 &  1 & 12 & 6.54\% & 6.54\% & 6.54\% \\
42 & AES Decrypt               & Import/Generate & InvalidKeyException                & System error         & 34 & Strongbox & 1 &  1 & 4,100  & 1.91\% & 6.40\% & 3.71\% \\
43 & AES Decrypt               & Import/Generate & IllegalBlockSizeException          & Code -22                      & 34 & TEE & 1 &  1 & 115 & 1.65\% & 2.39\% & 2.19\% \\
44 & AES Decrypt               & Import/Generate & IllegalBlockSizeException          & Code -65535   & 26 & TEE & 1 & 2 & 44 & 3.56\% & 5.69\% & 5.38\% \\
45 & AES Decrypt               & Generate        & IllegalBlockSizeException          & Invalid operation handle  & 30 & TEE & 1 &  1 & 11 & 1.14\% & 1.14\% & 1.14\% \\
46 & AES Decrypt               & Import/Generate & IllegalBlockSizeException          & System error   & 34 & Strongbox & 1 &  1 & 1,600 & 1.99\% & 3.00\% & 2.16\% \\
\textbf{47} & \textbf{HMAC}                      & \textbf{Import/Generate} & \textbf{InvalidKeyException}                & \textbf{Invalid key blob}     & \textbf{26, 27, 29} &  \textbf{TEE} & \textbf{1} &  \textbf{5} & \textbf{1,200} & \textbf{74.35\%} & \textbf{100\%} & \textbf{99.7\%} \\
48 & HMAC                      & Import/Generate & InvalidKeyException                & System error         & 34 &  Strongbox & 1 & 1 & 2,800 & 2\% & 5.8\% & 4.15\% \\
\textbf{49} & \textbf{ECDH Exchange}             & \textbf{Import}          & \textbf{InvalidKeyException}                & \textbf{Incompatible purpose} & \textbf{31-32} & \textbf{TEE/Software} & \textbf{10} & \textbf{146} & \textbf{3.8 million} & \textbf{43.661\%} & \textbf{100\%} & \textbf{100\%} \\
\textbf{50} & \textbf{ECDH Exchange}             & \textbf{Import}          & \textbf{InvalidKeyException}                & \textbf{Incompatible purpose} & \textbf{33} & \textbf{TEE/Software} & \textbf{7} & \textbf{116} & \textbf{4.2 million} & \textbf{41.79\%} & \textbf{100\%} & \textbf{100\%} \\
\textbf{51} & \textbf{ECDH Exchange}             & \textbf{Import}          & \textbf{InvalidKeyException}                & \textbf{Incompatible purpose} & \textbf{33} & \textbf{TEE/Software} & \textbf{7} & \textbf{122} & \textbf{6.4 million} & \textbf{40\%} & \textbf{100\%} & \textbf{100\%} \\
\bottomrule
\end{tabular}

}
\label{tab:operation_error}
\end{table*}

\begin{figure}[t]
     \centering
         \begin{subfigure}[b]{0.493\linewidth}
         \centering
         \includegraphics[width=\textwidth]{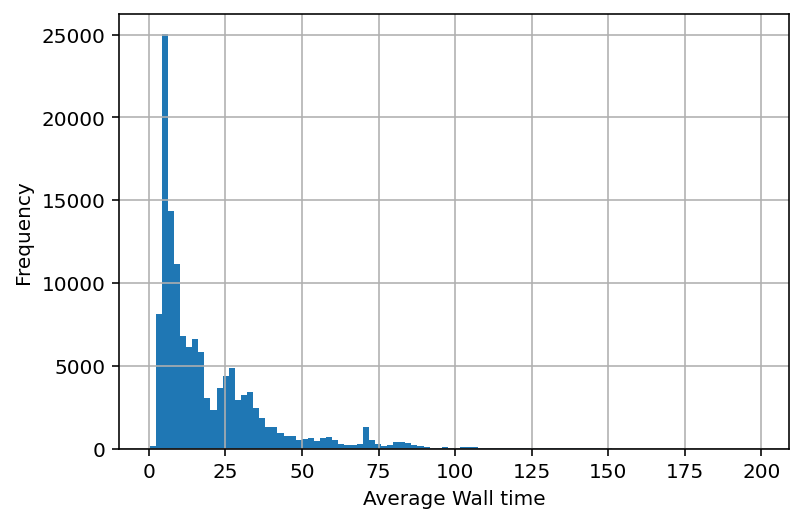}
         \caption{Distribution of wall time average for ECDSA signing for 130,000 devices.}
         \label{fig:ecdsa_wall_time_avg}
     \end{subfigure}
     \hfill
     \begin{subfigure}[b]{0.493\linewidth}
         \centering
         \includegraphics[width=\textwidth]{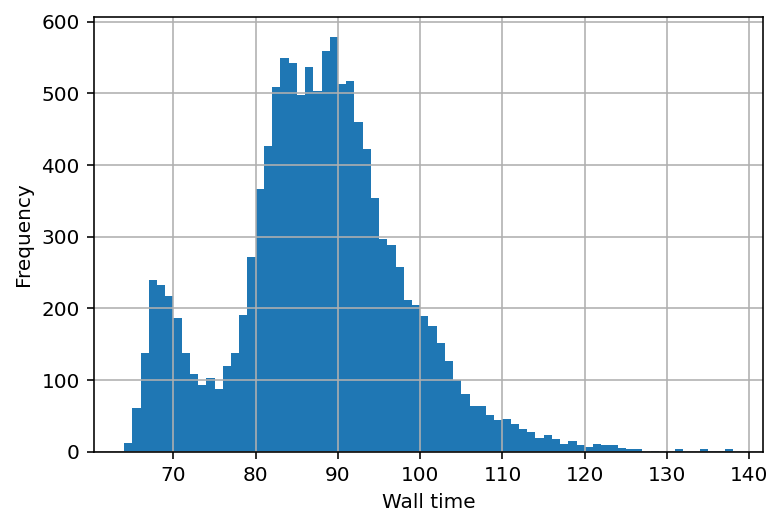}
         \caption{Distribution of ECDSA signing timing for a single device, StrongBox.}
         \label{fig:ecdsa_sample_timing}
     \end{subfigure}
    \caption{Average and sample ECDSA timing.}
\end{figure}

\end{document}